\documentclass{jfm}
\usepackage{graphicx}
\usepackage{subfig}
\usepackage{epstopdf, epsfig}
\usepackage{natbib}
\usepackage{amsmath,amssymb}

\usepackage{color}

\shorttitle{Kazantsev model in nonhelical $2.5D$ flows}
\shortauthor{K. Seshasayanan and A. Alexakis}

\title{Kazantsev model in nonhelical $2.5D$ flows}

\author{K. Seshasayanan\aff{1}\corresp{\email{skannabiran@lps.ens.fr}} and A. Alexakis\aff{1}}

\affiliation{\aff{1}Laboratoire de Physique Statistique, {\'E}cole Normale Sup{\'e}rieure, CNRS UMR 8550, Universit{\'e} Paris Diderot, Universit{\'e} Pierre et Marie Curie, 24 rue Lhomond, 75005 Paris, France}

\begin{document}

\maketitle

\begin{abstract}
We study the dynamo instability for a Kazantsev-Kraichnan flow with three velocity components that depends only on two-dimensions 
${\bf u}=\left( u(x, y, t), v(x, y, t), w(x, y, t) \right)$ often referred to as 2.5 dimensional ($2.5D$) flow. 
Within the Kazantsev-Kraichnan framework we derive the governing equations for the second order magnetic field correlation function
and examine the growth rate of the dynamo instability as a function of the control parameters of the system. 
In particular we investigate the dynamo behaviour for large magnetic Reynolds numbers $Rm$ and flows close to being two-dimensional and show that these two limiting procedures do not commute. 
The energy spectra of the unstable modes are derived analytically and lead to power-law behaviour that differs from the three dimensional and two dimensional case.
The results of our analytical calculation are compared with the results of numerical simulations of dynamos driven by prescribed fluctuating flows 
as well as freely evolving turbulent flows, showing good agreement.
\end{abstract}

\begin{keywords}
\end{keywords}

\section{Introduction}

Dynamo instability refers to the amplification of magnetic fields by the flow of a conducting fluid. It is responsible for the existence of magnetic fields in most astrophysical bodies. In most situations the driving flow is turbulent and this prevents an analytical treatment of the problem. Thus most studies are restricted to large scale numerical simulations or simplified models. A simple flow that can be treated analytically is the Kazantsev-Kraichnan flow. This model considers the kinematic dynamo instability driven by a random velocity field that is homogeneous, delta-correlated in time and gaussian distributed. It was first examined by \cite{kazantsev1968enhancement} for the dynamo instability and was independently studied by \cite{kraichnan1968small} for the problem of passive scalar advection. Physically, the delta-correlated time behaviour, models the fast varying turbulent scales of the velocity field. Under these assumption the problem can be simplified to a one dimensional eigenvalue problem, the eigenvalue of which gives the growth rate of the magnetic energy.

The Kazantsev-Kraichnan flow has been widely studied for three-dimensional  isotropic flows. 
Since the velocity field is gaussian distributed its statistics are entirely given by the second order correlation function. 
The correlation function $g^{ij} \left( {\bf r} \right)$ of the velocity field is defined as 
$ \left\langle u^i \left( {\bf x + r }, t \right) u^j \left( {\bf x}, t' \right) \right\rangle = g^{ij} \left( {\bf r} \right) \delta \left( t - t' \right)$ 
where due to homogeneity the function $g^{ij}$ is independent of ${\bf x}$. 
The first study by Kazantsev considered a 
flow for which the correlation function scales like $| g^{ii} \left( r \right) | \sim r^{\zeta}$ with $\zeta$ being the H\"older exponent. 
He found existence of dynamo instability in the range $1 < \zeta \leq 2$ for large $Rm$. 
Flows with H\"older exponents $\zeta < 2$ correspond to rough flows and model the turbulent scales
while flows with $\zeta = 2$ correspond to smooth velocity fields that model the viscous scales where the nonlinearities are in balance with the viscous dissipation.
Since then various authors \citep{ruzmaikin1981magnetic,novikov1983kinematic,falkovich2001particles,vincenzi2002kraichnan,schekochihin2002spectra}
have considered velocity fields with both a turbulent inertial range and a viscous scale cut-off at various limits of the system.
%
For smooth flows $\zeta = 2$, \cite{chertkov1999small} calculated the higher order moments and multipoint correlation functions by means of a Lagrangian approach. 
Geometric properties of the advected field were examined by \cite{boldyrev2000geometric} and the effect of  nonlinearities were examined in \cite{boldyrev2001solvable}.
More recently the predictions of the model as well as the non-linear behaviour have been examined by means of three-dimensional numerical simulations
\citep{schekochihin2004simulations,iskakov2007numerical,mason2011magnetic}. 

There is a major difference between a two dimensional ($2D$) flow and a three dimensional ($3D$) flow concerning the dynamo instability. 
$2D$ flows do not lead to a dynamo instability for any value of the magnetic Reynolds number as shown by \cite{zeldovich1957magnetic}. 
This is also true in the $2D$ Kazantsev model that has been examined in detail by \cite{schekochihin2002spectra}
and more recently the evolution of a $3D$ magnetic field by a $2D$ flow was examined by \cite{kolokolov2016kinematic}.
A careful analysis of the time evolving solution indicates that in two dimensions the energy of
any initial magnetic field localized in the wavenumber space will grow exponentially 
due to the increasing number of excited modes, even if the energy amplitude of each individual mode decreases. 
This behaviour persists until the length scale of the magnetic field becomes comparable to the dissipation scale after which dissipation becomes effective and
the total magnetic energy decays. 
The decaying magnetic field spectrum forms a power law behaviour with an exponent $k^2$. In contrast in the three dimensional case for sufficiently large $Rm$ an initial magnetic field localized in space has growing number of excited modes and each mode grows in time. The magnetic energy spectra in $3D$ has a powerlaw $k^{3/2}$ behaviour.

In this paper we are interested in developing the Kazantsev model for a flow where the velocity field takes the form 
${\bf u} = \left( u(x, y, t), v(x, y, t), w(x, y, t) \right)$, meaning it has three components but depends only on two-dimensions.
Such flows are refered in the literature as $2.5D$ flows. 
They can be considered as the limiting case of a very fast rotating system for which,
according to the Taylor-Proudmann theorem \citep{proudman1916motion, taylor1917motion}, the flow
becomes two-dimensional due to the Coriolis force that suppresses fluctuations along the direction of rotation. 
$2.5D$ flows are some of the simplest flows that give rise to the dynamo instability and have been extensively studied 
for smoothly varying flows \citep{roberts1972dynamo,galloway1992numerical}. 
%
Our interest lies on turbulent flows that have been examined recently at various contexts \cite{smith2004vortex, tobias2008dynamo,seshasayanan2015turbulent}
where the dynamo instability driven by a turbulent $2.5D$ flow has been studied in detail. 
In \cite{seshasayanan2015turbulent} it was shown that both helical and non-helical 2.5D flows can lead to a dynamo instability. 
For the helical flow and for small $Rm$ the instability can be explained by an $\alpha$-effect. 
The $\alpha$-effect is a mean field effect where the small scale magnetic field and the small scale velocity field interact to amplify the magnetic fields at large scales. 
For the non-helical flow however the $\alpha$-coefficient is zero and it does not provide an explanation for the observed dynamo growth rates.
Thus this dynamo remains theoretically unexplained.

The main purpose of this work is to examine analytically the dynamo instability for the nonhelical flow for the Kazantsev-Kraichnan model for the $2.5D$ flow. 
We first derive a system of equations that govern the second order correlation function of the magnetic field. 
This leads to a linear system of equations and an eigenvalue problem which is then solved for a model velocity field that we consider. 
This allows us to explicitly calculate the growthrate and the spectral behaviour of the most unstable modes.
We restrict to the case of smooth, velocity fields with a correlation function that scales like $r^2$ at small scales. 

The rest of the article is constructed in the following way. 
Section \ref{Section:Two} describes the governing equations on which this study is based. 
We set-up a model flow to be studied in section \ref{Section:Three}. 
The dynamo instability properties of this model flow is examined in section \ref{Section:Four} and in \ref{Section:Five}. 
Section \ref{Section:Six} describes the spectral behaviour of the most unstable eigen-mode. 
In section \ref{Section:Seven}, we compare the analytical results with the results from numerical simulations. 
Finally in section \ref{Section:Eight} we conclude the study and give some future perspectives.
 
\section{The model} \label{Section:Two}

We consider a $2.5D$ flow of the form ${\bf u}(x, y, t) = (u_x, u_y, u_z)$ which can also be written in terms of the stream function 
$\psi (x, y)$ as ${\bf u} = \nabla \times (\psi \hat{\bf e}_z) + u_z \hat{\bf e}_z = {\bf u}_{_{2D}} + u_z \hat{\bf e}_z$ 
where $z$ is the invariant direction. The Kazantsev-Kraichnan ansatz considers the velocity field to be delta correlated in time, gaussian distributed, its statistics is entirely governed by the second order correlation function. We further consider that the velocity field is homogeneous and $2D$ isotropic in the plane $x, y$. Isotropy in $2D$ means that the statistics of the velocity field is invariant under rotations around the $z$-axis. The correlation function of two components of the velocity field $u^i, u^j$ at points ${\bf x + r, x}$ can be written as,
\begin{eqnarray}
\left\langle u^i \left( {\bf x + r}, t \right) u^j \left( {\bf x}, t' \right) \right\rangle = g^{ij} \left( {\bf r} \right) \delta \left( t - t' \right).
\end{eqnarray}
Independence of $g^{ij}$ on ${\bf x}$ emerges from homogeneity.

The general form of an isotropic second order correlation function $g^{ij} \left( {\bf r} \right)$ for a $2.5D$ flow (see \cite{oughton1997general}) is given by,
\begin{align} 
g^{ij}\left( {\bf r} \right) = & g_{_{LL}} \left( r \right) \delta^{ij} - \Big( g_{_{LL}} - g_{_{NN}} \Big) \left( \delta^{ij} - \frac{r^i r^j}{r^2} \right) + \left( g_{_{Z}} \left( r \right) - g_{_{2D}} \left( r \right) - g_{_{2D}}' \left( r \right) r \right) \delta^{i3} \delta^{j3} \nonumber \\
& + g_c \left( r \right) \Big( \delta^{i3} \frac{r^j}{r} - \frac{r^i}{r} \delta^{j3} \Big) + g_p \left( r \right) \Big( \epsilon^{3jp} \delta^{i3} \frac{r^p}{r} - \epsilon^{3ip} \delta^{j3} \frac{r^p}{r} \Big) \label{eqn:velstructfns}
\end{align} 
where $\delta^{ij}$ is the Kronecker delta tensor and $\epsilon^{ijk}$ is the Levi-Civita tensor. The indices $i, j$ take the values $1, 2, 3$. All the quantities depend only on two-dimensions in space, hence we have used a projected coordinate ${\bf r} = (x,y,0) = (r^1, r^2, r^3)$ in equation \ref{eqn:velstructfns}. The derivative of $g^{ij}\left({\bf r}\right)$ with respect to $r^3 = z$ is zero. The prime on a scalar function $g'$ denotes the derivative with respect to $r$. The functions $g_{_{LL}}, g_{_{NN}}, g_c, g_{p}, g_{_{Z}}$ are scalar functions that depend only on $r$ and are defined as,
\begin{align}
 g_{_{LL}} \left( r \right) = \left\langle ({\bf e}_r \cdot {\bf u})( {\bf u}' \cdot  {\bf e}_r)                     \right\rangle_T&,  \quad
 g_{_{Z}}  \left( r \right) = \left\langle ({\bf e}_z \cdot {\bf u})( {\bf u}' \cdot  {\bf e}_z)                     \right\rangle_T,  \nonumber \\
 g_c       \left( r \right) = \left\langle ({\bf e}_z \cdot {\bf u})( {\bf u}' \cdot  {\bf e}_r)                     \right\rangle_T&,  \quad
 g_p       \left( r \right) = \left\langle ({\bf e}_z \cdot {\bf u})( {\bf u}' \cdot ({\bf e}_z \times \hat{\bf r})) \right\rangle_T,    \\ 
 g_{_{NN}} \left( r \right) = \left\langle (({\bf e}_z \times{\bf e}_r) \cdot {\bf u})({\bf u'} \cdot ({\bf e}_z \times {\bf e}_r))\right\rangle_T  &,  \nonumber
\end{align}
where $\hat{\bf r}$ is the unit vector along ${\bf r}$ direction. ${\bf u}$ is the velocity field at a point ${\bf x + r}$ at time $t$, ${\bf u'}$ is the velocity field at a point ${\bf x}$ at time $t'$, the symbol $\left\langle \; \; \right\rangle_T$ denotes both time average and ensemble average. Physically the quantity $g_{_{LL}}$ measures the longitudinal auto correlation function of the two-dimensional velocity field. The quantity $g_{_{NN}}$ gives the transverse auto correlation of the two dimensional velocity field. $g_c$ and $g_p$ are the cross correlation between the two-dimensional velocity field and the vertical velocity field. The function $g_{_{Z}}$ gives the autocorrelation of the vertical velocity field. In particular the function $g_p$ is related to the helicity of the velocity field. Since we consider a velocity field that is nonhelical, we take $g_p \left( r \right) = 0$. The incompressibility condition for the velocity field $\partial_x u_x + \partial_y u_y = 0$ implies for the correlation function, $g^{ij}_{,i} = g^{ij}_{,j} = 0$, where the subscript ${}_{,i}$ in ${g^{ij}_{,i}}$ denotes differentiation of $g^{ij}$ with respect to $r^i$. This implies,
\begin{align}
g_{_{NN}} \left( r \right) & = g_{_{LL}} \left( r \right) + g'_{_{LL}} \left( r \right) r \\
g_c \left( r \right) & = 0
\end{align}
leaving two functions $g_{_{LL}} \left( r \right), g_{_{Z}} \left( r \right)$ that determine fully the second order velocity correlation function. 

Due to the invariance of the velocity field along $z$-direction the perturbations of the magnetic field can be decomposed into Fourier modes of the form ${\bf B} = {\bf b}(x, y, t) \, exp(i k_z z)$. The complex vector field ${\bf b}$ is governed by the induction equation which can be written as, 
\begin{eqnarray}
\partial_t {\bf b} + \left( \nabla \times \psi \hat{\bf e}_z \right) \cdot \nabla \, {\bf b} + u_z i k_z {\bf b} = {\bf b} \cdot \nabla \, \left( \nabla \times \psi \hat{e}_z \right) + \eta \, \left( \Delta - k_z^2 \right) {\bf b} \label{eqn:induction}
\end{eqnarray} 
where $\eta$ is the magnetic diffusivity. The solenoidal condition for the magnetic field $\nabla \cdot {\bf B} = 0$ gives, 
\begin{eqnarray}
\partial_x b_x(x, y, t) + \partial_y b_y(x, y, t) = - i k_z b_z(x, y, t) \label{eqn:solenfirst}
\end{eqnarray}
where ${\bf b} = (b_x, b_y, b_z)$. The evolution of the magnetic field can be quantified by considering the second order correlation function defined as,
\begin{eqnarray}
H^{ij} \left( {\bf r}, t \right) = \left\langle \left( b^i \left( {\bf x + r, t}\right) \right)^{\dagger} b^j \left( {\bf x, t} \right) \right\rangle
\end{eqnarray}
where the symbol ${}^{\dagger}$ denotes the complex conjugate. As shown in the appendix \ref{Appendix:One}, given that the velocity field is mirror symmetric and the governing equation is of the form equation \ref{eqn:induction} we only need to look at the mirror symmetric part of the magnetic field. This is because the induction equation in the absence of a mirror asymmetric part in the velocity field leads to a decoupled equation for the mirror symmetric and the mirror asymmetric part. Thus we only need to concentrate on the mirror symmetric part of the magnetic field neglecting magnetic helicity similar to most studies of Kazantsev model in $3D$, see however \cite{subramanian1999unified, boldyrev2005magnetic}, where a helical flow is considered and the magnetic helicity is present. The general form of the magnetic correlation function for a nonhelical complex field can be written as, 
\begin{eqnarray}
H^{ij} \left( {\bf r}, t \right) = & H_{_{LL}} \left( r \right) \delta^{ij} - \Big( H_{_{LL}} \left( r \right) - H_{_{NN}} \left( r \right) \Big) \Big( \delta^{ij} - \frac{r^i r^j}{r^2} \Big) + \Big( H_{_{Z}} \left( r \right) - H_{_{NN}} \left( r \right) \Big) \delta^{i3} \delta^{j3} \nonumber \\
& + i \; {{H}}_c \left( r \right) \Big( \delta^{i3} \frac{r^j}{r} + \frac{r^i}{r} \delta^{j3} \Big). \label{eqn:genformH}
\end{eqnarray}
where $H_{_{LL}}, H_{_{NN}}, {H}_c, H_{_{Z}}$ are scalar real functions that only depend on $r$ and are defined as,
\begin{align}
H_{_{LL}} \left( r, t \right) &=  \left\langle  ({\bf e}_r \cdot {\bf b}^{\dagger})( {\bf b'} \cdot {\bf e}_r) \right\rangle_T , \quad
&H_c \left( r, t \right)       &= \left\langle  ({\bf e}_z \cdot {\bf b}^{\dagger})( {\bf b} \cdot  {\bf e}_r  \right\rangle_T, \nonumber \\
H_{_{NN}} \left( r, t \right) &=  \left\langle (({\bf e}_z \times {\bf e}_r) \cdot {\bf b}^{\dagger} )({\bf b} \cdot ({\bf e}_z \times {\bf e}_r))\right\rangle_T, \quad
&H_{_{Z}} \left( r, t \right)  &= \left\langle  ({\bf e}_z \cdot {\bf b}^{\dagger})( {\bf b} \cdot {\bf e}_z) \right\rangle_T.
\end{align}
where ${\bf b}$ is the magnetic field at a point ${\bf x + r}$ at time $t$ and ${\bf b'}$ is the magnetic field at a point ${\bf x}$ at time $t$. This general form can be derived by writing the magnetic field in terms of scalar functions and then writing the two point correlation function in terms of these scalar functions (see \cite{oughton1997general}). The function $H_{_{LL}}$ is the longitudinal auto correlation function of the two dimensional magnetic field and $H_{_{NN}}$ is the transverse auto correlation function of the two-dimensional magnetic field. The function ${H}_c$ is the cross correlation function of the two-dimensional magnetic field with the vertical magnetic field $b_z$. $H_{_{Z}}$ is the auto-correlation function of vertical magnetic field $b_z$. The solenoidal condition of the magnetic field (equation \ref{eqn:solenfirst}) for the correlation function implies,
\begin{align}
H^{ij}_{,i} - i k_z H^{3j} = 0, \hspace{10mm} H^{ij}_{,j} - i k_z H^{i3} = 0 
\end{align}
which gives the set of following relations for the scalar correlation functions, 
\begin{align}
k_z H_{_{Z}} \left( {\bf r} \right) & = {H}_c' \left( {\bf r} \right) + \frac{{H}_c \left( {\bf r} \right)}{r} \label{eqn:incomfirst} \\
- k_z {H}_c \left( {\bf r} \right) & = H_{_{LL}}' \left( {\bf r} \right) + \frac{H_{_{LL}} \left( {\bf r} \right) - H_{_{NN}} \left( {\bf r} \right)}{r} \label{eqn:incomsecond}
\end{align} 
When $k_z = 0$ we get ${H}_c = 0$ and $H_{_{NN}} = H_{_{LL}} + r H'_{_{LL}}$. If the magnetic field is $2.5D$, the magnetic correlation function $H^{ij}$ becomes real and it simplifies to a form similar to the velocity correlation function $g^{ij}$.

Given the velocity correlation functions $g^{ij}$ it is possible to derive the governing equation for $H^{ij}$ starting from the induction equation \ref{eqn:induction}. The governing equation for $H^{ij}$ leads to triple product correlations of velocity and magnetic fields. The triple product can be written in terms of second order correlation functions of the velocity and the magnetic field by using the Furutsu-Novikov theorem \citep{furutsu1963nbs, novikov1965functionals}. This theorem uses the fact that the velocity field is Gaussian distributed.  
Due to the solenoidal conditions (equation \ref{eqn:incomfirst}, \ref{eqn:incomsecond}) only two equations are required to completely determine the magnetic correlation function $H^{ij}$ that we here chose to be $H_{_{LL}}, {H}_c$. The governing equations then read
\begin{align}
\partial_t H_{_{LL}} - & \Big( 2 \eta + g_{_{LL}} \left( 0 \right) - g_{_{LL}} \Big) \Big[ H''_{_{LL}} + 3 \frac{H'_{_{LL}}}{r} \Big] + k_z^2 \Big( 2 \eta + g_{_{Z}} \left( 0 \right) - g_{_{Z}} \Big) H_{_{LL}} = - g''_{_{LL}} H_{_{LL}} \nonumber \\
& - g'_{_{LL}} \Big( 2 H'_{_{LL}} + 3 \frac{H_{_{LL}}}{r} \Big) - 3 k_z {H}_{c} \; g'_{_{LL}} + \frac{2}{r} \Big( 2 \eta + g_{_{LL}} \left( 0 \right) - g_{_{LL}} \Big) k_z {H}_{c} \\
\partial_t {H}_{c} - & \Big( 2 \eta + g_{_{LL}} \left( 0 \right) - g_{_{LL}} \Big) \Big[ {H}''_{c} + \frac{1}{r} {H}'_{c} - \frac{1}{r^2} {H}_{c} \Big] + k_z^2 \Big( 2 \eta + g_{_{Z}} \left( 0 \right) - g_{_{Z}} \Big) {H}_{c} = - k_z g'_{_{Z}} H_{_{LL}}.  \label{eqn:magcorreqns}
\end{align}
The details of the derivation are given in the Appendix \ref{Appendix:One}.
The quantity $g_{_{LL}} \left( 0 \right)$ is the total energy of the velocity field in $2D$ while the quantity $g_{_{Z}} \left( 0 \right)$ is the total energy of the velocity in the $z$ direction. These terms, $g_{_{LL}}(0),g_{_{Z}}(0)$, depend on the frame of reference from which they are measured and do not modify the dynamo instability. 

We identify three special cases which do not lead to a dynamo instability. 
\begin{enumerate}
\item When $k_z = 0$ the equations simplify to the $2D$ Kazantsev model which does not give rise to the dynamo instability as shown in previous studies (see for example \cite{schekochihin2002spectra}). This means that $k_z \ne 0$ is required in order to have a dynamo instability.
\item When the third velocity component is zero $u_z = 0$ then $g_{_{Z}} = 0$. This leads to the function ${H}_c$ no longer being driven/coupled to $H_{_{LL}}$. In the presence of diffusivity in the long time limit ${H}_c$ would decay to zero. Alternatively we can show that the governing equation for the vertical magnetic field is an advection-diffusion equation without any forcing. Thus the vertical magnetic field $b_z$ decays in the long time limit. In the absence of ${H}_c$ the equations governing $H_{_{LL}}$ become again the $2D$ Kazantsev equations and hence $H_{_{LL}}$ would also decay in the long time limit.
\item The case when there is no shear in the two dimensional flow $g_{_{LL}} = g_{_{LL}} \left( 0 \right)$ does not lead to a dynamo instability. The component $b_z$ can be amplified by the stretching of $b_x, b_y$ by $u_z$. But it can be seen from the induction equation that the magnetic fields components $b_x, b_y$ are advected by $u_z$ and dissipated by the ohmic dissipation with no amplification from the stretching term. Thus both $b_x, b_y$ decay in the long time limit which makes $b_z$ to decay in the long time limit.
These special cases fall under the Zeldovich anti-dynamo theorem for $2D$ flows. Hence the velocity field has to have all the three components and $k_z \neq 0$ in order for the existence of the dynamo instability in the long time limit. 
\end{enumerate}

In the next section we will consider a model flow where we calculate the form for the functions $g_{_{LL}} \left( r \right), g_{_{Z}} \left( r \right)$. 
We then proceed to study the dynamo instability driven by this model flow in terms of the other control parameters of the system. 

\section{Model flow} \label{Section:Three}

We consider a smooth isotropic and homogeneous velocity field given in terms of the stream function $\psi$ and the vertical velocity $u_z$ as,
\begin{align}
\psi \left( {\bf r}, t \right) = & \zeta_1 \left( t \right) \sin \Bigg( \frac{k_0}{2} \Big[ \sin \left( \phi_1 \left( t \right) \right) x + \cos \left( \phi_1 \left( t \right) \right) y \Big] + \phi_2 \left( t \right) \Bigg) \\
u_z \left( {\bf r}, t \right)  = & \zeta_2 \left( t \right) \cos \Bigg( \frac{k_0}{2} \Big[ \sin \left( \phi_1 \left( t \right) \right) x + \cos \left( \phi_1 \left( t \right) \right) y \Big] + \phi_2 \left( t \right) \Bigg).
\end{align} \label{eqn:modelflow}
$\phi_1 \left( t \right), \phi_2 \left( t \right)$ are random variables which are uniformly distributed over $[0, 2 \pi]$ and render the flow homogeneous and isotropic. 
$\zeta_1 \left( t \right)$ and $\zeta_2 \left( t \right)$ are random variables that are Gaussian distributed in time with 
$\left\langle \zeta_1 \left( t \right) \zeta \left( t' \right) \right\rangle = \Theta_1 \delta \left( t - t' \right)$, 
$\left\langle \zeta_2 \left( t \right) \zeta_2 \left( t' \right) \right\rangle = \Theta_2 \delta \left( t - t' \right)$ and
$\left\langle \zeta_1 \left( t \right) \zeta_2 \left( t' \right) \right\rangle =0$.
The wavenumber $k_0$ defines a typical length scale for the velocity field. 
The correlation function of the velocity field is calculated to be,
\begin{align}
g^{ij} \left( {\bf r} \right) = \frac{k_0 \Theta_1}{4} & \Bigg\{ -\delta^{ij} \frac{J'_0 \left( k_0 \frac{r}{2} \right)}{r} + \Big( \delta^{ij} - \frac{r^i r^j}{r^2} \Big) \Big( \frac{J'_0 \left( k_0 \frac{r}{2} \right) }{r} - \frac{k_0}{2} J''_0 \left( k_0 \frac{r}{2} \right) \Big) \Bigg\} \nonumber \\
+ & \frac{\Theta_2}{2} J_0 \left( k_0 \frac{r}{2} \right) \,\, \delta^{i3}\delta^{j3}
\end{align}
where $J_0$ is the Bessel function of the first kind and $J'_0$ stands for its derivative. The functions $g_{_{2D}}, g_{_{Z}}$ are then,
\begin{align}
g_{_{2D}} \left( r \right)  = - \frac{k_0 \Theta_1}{4r} J'_0 \Big( k_0 \frac{r}{2} \Big), \hspace{10mm}
g_{_{Z}} \left( r \right) = \frac{\Theta_2}{2} J_0 \Big( k_0 \frac{r}{2} \Big)
\end{align}
The small $r$ behaviour of these functions is,
\begin{align}
g_{_{2D}} \left( r \right) = g_{_{2D}} \left( 0 \right) - D_1 r^2 + E_1 r^4 - O \left( r^6 \right), \hspace{7mm} g_{_{Z}} \left( r \right) = g_{_{Z}} \left( 0 \right) - D_2 r^2 + E_2 r^4 - O \left( r^6 \right) 
\end{align}
where $g_{_{2D}} \left( 0 \right) = k_0^2 \Theta_1/16, g_{_{Z}} \left( 0 \right) = \Theta_2/2, D_1 = k_0^4 \Theta_1/512, D_2 = k_0^2 \Theta_2/32$. At small scales the velocity field is smooth and behaves like $g_{_{2D}} \sim r^2, g_{_{Z}} \sim r^2$. 

 We note that the $D_1$ has dimensions of inverse time and defines the dynamical time scale $\tau_d \equiv 1/D_1$ that we will use to non-dimensionalize our system.
Accordingly the magnetic Reynolds number is defined as the ratio of 
the diffusion time scale $1/\eta k_0^2$ to the dynamical time scale 
$Rm \equiv D_1/(k_0^2 \eta) = k_d^2/k_0^2$ where $k_d$ is the dissipation length scale for the magnetic field 
$k_d \equiv k_0 \sqrt{D_1/\eta} = k_0 \sqrt{Rm}$. 
%
 A third dimensionless parameter can be defined by the ratio 
of the vertical velocity field gradients to the planar velocity field gradients 
the we will quantify as $D_r = D_2/D_1$. The quantity $D_r$ depends on the ratio of the amplitudes of $k_0^2\Theta_1$ and $\Theta_2$ 
given in equation \ref{eqn:modelflow} as $D_r = 16 \Theta_2/\left( \Theta_1 k_0^2 \right)$.
Thus the nondimensionalized control parameters are, the wavemode $k_z/k_0$, the magnetic Reynolds number $Rm$ and $D_r$.

\section{Growth rate $\gamma$} \label{Section:Four}
\begin{figure}
\begin{center}
\includegraphics[scale=0.2]{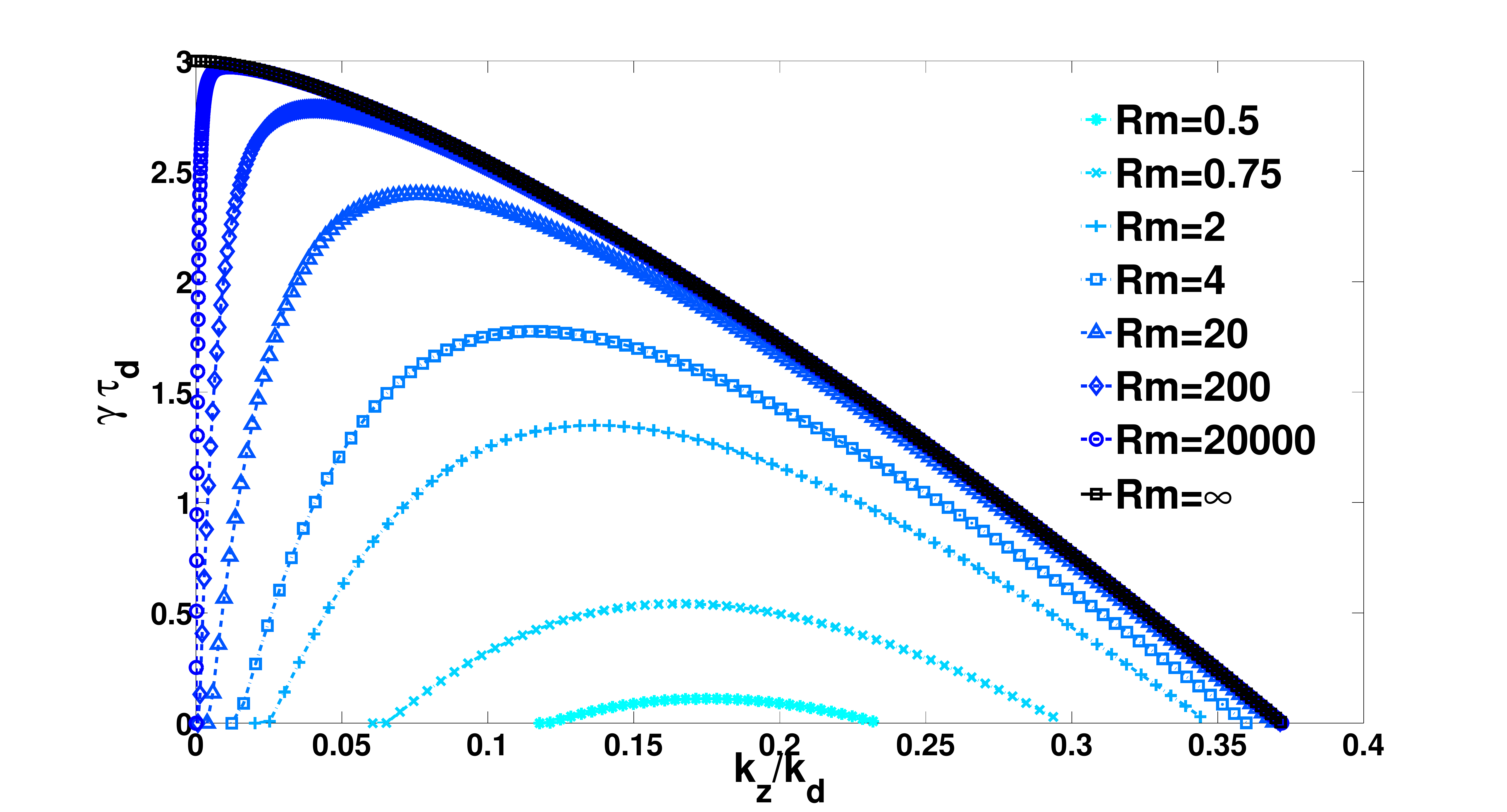}
\end{center}
\caption{Normalized growth rate $\gamma \tau_d$ is shown as a function of the normalized modes $k_z/k_d$ for different values of $Rm$. Darker shades correspond to larger values of $Rm$.}
\label{fig:Growth1}
\end{figure}

Substituting $H_{_{LL}} = e^{\gamma t} h_{_{LL}}$ and $H_c = e^{\gamma t} h_c$ in equation \ref{eqn:magcorreqns} 
we end up with an eigenvalue problem for the growth rate of the magnetic energy $\gamma$ and the eigenfunctions $h_{_{LL}}$ and $h_c$.
The largest eigenvalue of the system $\gamma$ controls the long time evolution of the magnetic field correlation functions.
We note that since $H_{_{LL}}$ and $ H_c$ are quadratic quantities in the magnetic field $\bf b$ the growth rate $\gamma$ is twice  
the growth rate of the magnetic field.
We proceed in this section by solving the resulting system of equations numerically. 
To solve the eigenvalue problem we use a Chebyshev spectral method to discretize the domain, and we project the functions $h_{_{LL}} \left( r \right), h_c \left( r \right), g_{_{LL}} \left( r \right), g_{_{Z}} \left( r \right)$ into a truncated basis of Chebyshev functions. The equations \ref{eqn:magcorreqns} in this truncated basis can now be reduced to a linear matrix eigenvalue problem. We compute the largest positive eigenvalue of the discretized matrix using standard linear algebra software. We have checked the convergence of the resulting eigenvalue in terms of the number of basis functions used and the domain size taken.

Figure \ref{fig:Growth1} shows the growth rate $\gamma$ as a function of the rescaled parameter $k_z/k_d$ for different values of $Rm$. Dynamo instability appears at values of $Rm$ above the critical magnetic Reynolds number $Rm_c$ which is found to be $Rm_c \approx 0.45$. Close to $Rm_c$ the instability occurs at the value $k_z \approx 0.18 k_d \approx 0.12 k_0$. For larger values of $Rm$ the instability is found in a range of wavenumbers $k_{min} < k < k_{max}$. The maximum value of $k_z/k_d$ at which the dynamo instability occurs initially increases with $Rm$ but reaches a constant value independent of $Rm$ for large value of $Rm$. We remind that $k_d \propto k_0 \sqrt{Rm}$ thus the largest wavenumber $k_{max}$ for which there is a dynamo instability increases like 
$k_{max} \sim k_0 \sqrt{Rm}$. The smallest wavenumber at which dynamo instability occurs $k_{min}$ decreases as we increase $Rm$. 
The growth rate of each mode $k_z$ increases as we increase $Rm$ reaching an asymptotic value at large $Rm$. 
The supreme of the growth rate $\gamma \tau_d =3$ is obtained for $Rm \rightarrow \infty$ and $k_z \rightarrow 0$. 
For very large $Rm$ we see that the curves themselves seem to reach an asymptotic behaviour which is captured well 
by the black solid curve representing the growth rate in the limit of $Rm \rightarrow \infty$ that we discuss in the next section.

\section{Three limiting behaviour} \label{Section:Five}

In this section we look at three different limits of the control parameters. 

\subsection{$Rm\rightarrow \infty, \quad k_z \rightarrow 0$} \label{Subsec:kzzero}

The limit of very large $Rm$ can be taken by letting the quantity $\eta \rightarrow 0$ in the equations \ref{eqn:magcorreqns}. In this limiting procedure we do the following change of variables, $\tilde{r} = r \, k_d$, $\tilde{t} = t\eta k_d^2 = t/D_1$. The velocity correlation functions are expanded in the following way, $g_{_{2D}} \left( \tilde{r} \right)= g_{_{2D}} \left( 0 \right) - \eta \tilde{r}^2 + O(\eta^2 \tilde{r}^4), g_{_{Z}} = g_{_{Z}} \left( 0 \right) - D_r \eta \tilde{r}^2 + O(\eta^2 \tilde{r}^4)$. Simplifying the resulting equation by considering only the highest order term we get,  
\begin{align}
\gamma \tau_d h_{_{LL}} & - \Big( 2 + \tilde{r}^2 \Big) \Big[ h''_{_{LL}} + 3 \frac{h'_{_{LL}}}{\tilde{r}} \Big] + \tilde{k}_z^2 \Big( 2 + D_r \tilde{r}^2 \Big) h_{_{LL}} = 2 h_{_{LL}} \nonumber \\
& 2 \tilde{r} \Big( 2 h'_{_{LL}} + 3 \frac{h_{_{LL}}}{\tilde{r}} \Big) + 6 \tilde{r} \tilde{k}_z {h}_{c} \; + \frac{2}{\tilde{r}} \Big( 2 + \tilde{r}^2 \Big) \tilde{k}_z {h}_{c}, \\
\gamma \tau_d {h}_{c} & - \Big( 2 + \tilde{r}^2 \Big) \Big[ {h}''_{c} + \frac{1}{\tilde{r}} {h}'_{c} - \frac{1}{\tilde{r}^2} {h}_{c} \Big] + \tilde{k}_z^2 \Big( 2 + D_r \tilde{r}^2 \Big) {h}_{c} = 2 \tilde{k}_z D_r \tilde{r} h_{_{LL}}. \label{eqn:magcorrinf}
\end{align}
The growth rate in this limit does not depend on $k_0$ but only on the local structure of the velocity field described by $D_r$. The eigenvalues of the black solid curve in figure \ref{fig:Growth1} were obtained by solving the above set of equations. It is important to note that the above set of equations are obtained in the limit of $\eta \rightarrow 0$ and not the case of $\eta = 0$. We find that the value of $\gamma \left( k_z \rightarrow 0 \right) = 3$ as $Rm \rightarrow \infty$. This value can be obtained by a matched asymptotic expansion that is described in section \ref{Section:Six} and in Appendix \ref{Appendix:Four}. On the other hand for a finite $Rm$ we see that $\gamma \left( k_z \rightarrow 0 \right) = 0$. Thus we have the non-commuting limits,
\begin{align}
	3 = \lim_{k_z \rightarrow 0} \lim_{Rm \rightarrow \infty} \gamma \neq \lim_{Rm \rightarrow \infty} \lim_{k_z \rightarrow 0} \gamma = 0. \label{eqn:firstnoncomm}
\end{align}
We mention here that the anti-dynamo theorem is still respected since it corresponds to the second limiting procedure above. 

\subsection{$Rm\rightarrow \infty, D_r \rightarrow 0$}
\begin{figure}
\begin{center}
\includegraphics[scale=0.2]{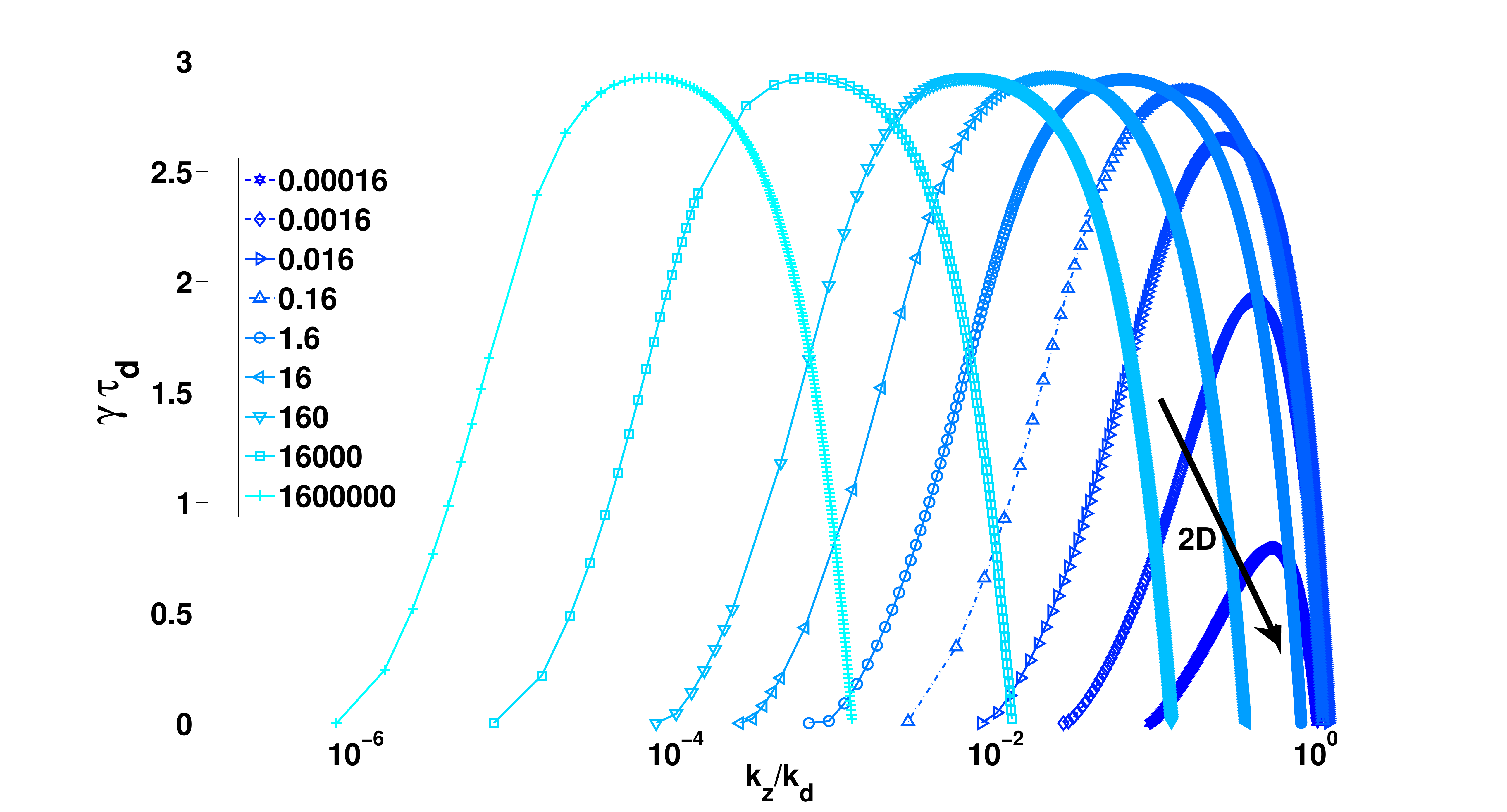}
\includegraphics[scale=0.2]{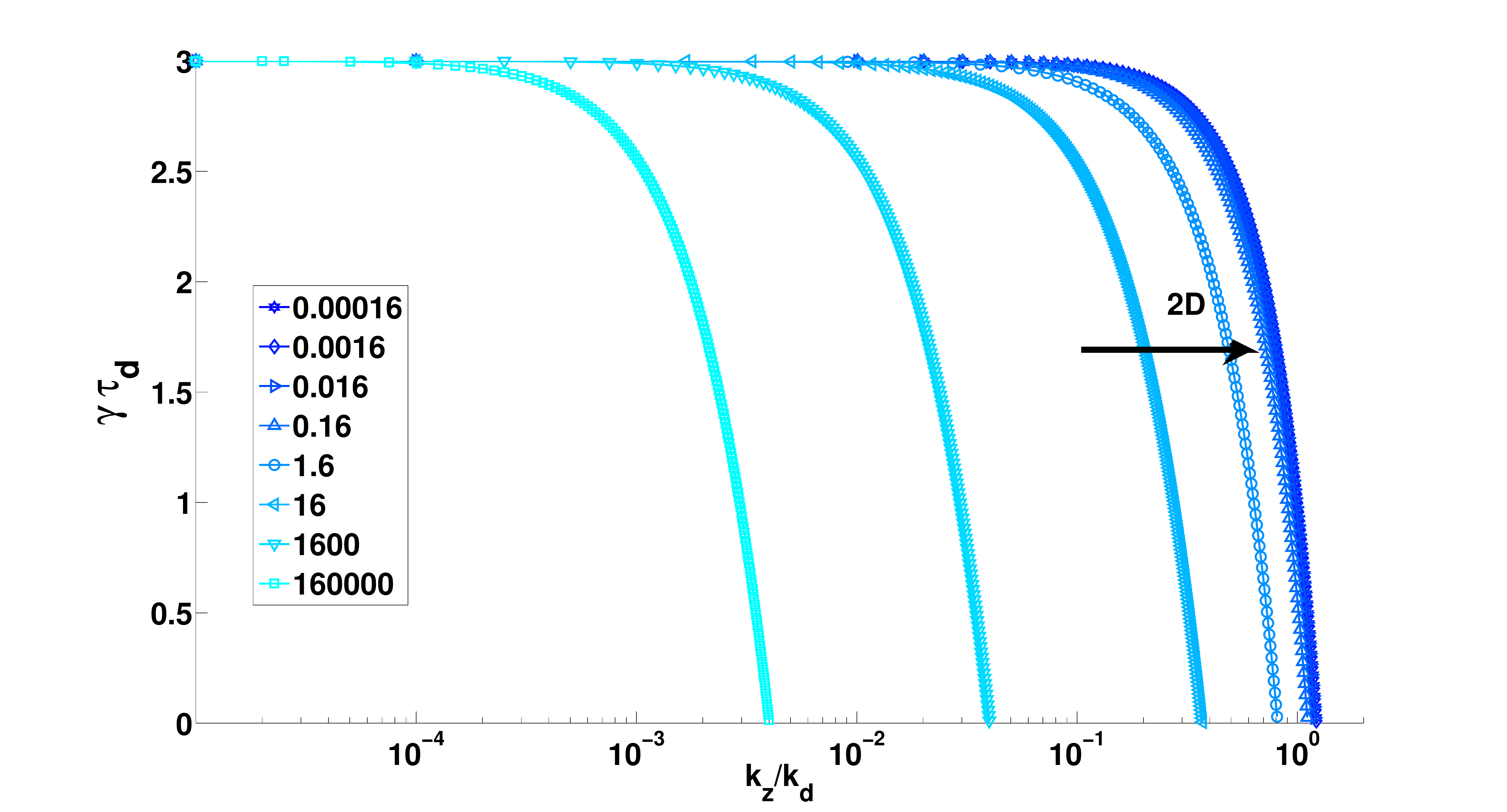}
\end{center}
\caption{Normalized growth rate $\gamma$ as a function of $k_z/k_d$ for different values of $D_r$ mentioned in the legends for, top - a finite $Rm \approx 1.95\mathrm{e}{+05}$, bottom - the case $Rm \rightarrow \infty$. The black arrow marked $2D$ shows the direction of decreasing value of the parameter $D_r$. Darker shades correspond to smaller values of $D_r$.}
\label{fig:becom2d}
\end{figure}

Taking the limit $D_r \rightarrow 0$ reduces the flow to a two-dimensional flow and from the anti-dynamo theorem we expect the dynamo instability to disappear. 
In figure \ref{fig:becom2d} we show $\gamma$ as a function of $k_z/k_d$ for a finite $Rm$ case on the top and for the case of $Rm \rightarrow \infty$ on the bottom 
for different values of the parameter $D_r$ as mentioned in the respective legends. The growth rate $\gamma$ and the range of unstable modes $k_z$ depend on the value of $D_r$. 
In the top panel of figure \ref{fig:becom2d} we see that indeed for the finite $Rm$ case as $D_r$ is decreased the dynamo instability disappears. 
This limit is pointed out in the plot by the arrow marked $2D$. 
On the contrary
for the case of the $Rm \rightarrow \infty$ (see bottom panel of figure \ref{fig:becom2d}) the growth rate $\gamma$ curve reaches a non-zero asymptotic 
behaviour as $D_r \rightarrow 0$ marked in the figure by the arrow 2D. 
Thus we obtain another set of non-commuting limits,
\begin{align}
0 < \lim_{D_r \rightarrow 0} \lim_{Rm \rightarrow \infty} \gamma \neq \lim_{Rm \rightarrow \infty} \lim_{D_r \rightarrow 0} \gamma = 0
\end{align}

This result needs to be explained. 
The case of $D_r = 0$ is a purely $2D$ flow and does not give rise to the dynamo instability in accordance with the anti-dynamo theorem which is respected by the governing equations. 
We can capture the limit of $D_r \rightarrow 0$ taken after the limit $Rm \rightarrow \infty$ 
by applying the following rescaling, $\sqrt{D_r} \tilde{r} \rightarrow \tilde{\tilde{r}}$, $h_c \rightarrow \sqrt{D_r} \tilde{h}_c$ to equations \ref{eqn:magcorrinf}. 
The highest order which captures the limit $D_r \rightarrow 0$ leads to the following set of equations,
\begin{eqnarray}
\gamma \tau_d h_{_{LL}} - \tilde{\tilde{r}}^2 \Big[ h''_{_{LL}} + 3 \frac{h'_{_{LL}}}{\tilde{\tilde{r}}} \Big] + k_z^2 \Big( 2 + \tilde{\tilde{r}}^2 \Big) h_{_{LL}} = & 2 h_{_{LL}} \\
2 \tilde{\tilde{r}} \Big( 2 h'_{_{LL}} + 3 \frac{h_{_{LL}}}{\tilde{\tilde{r}}} \Big) + & 8 \tilde{\tilde{r}} k_z \tilde{h}_{c} \nonumber \\
\gamma \tau_d \tilde{h}_{c} - \tilde{\tilde{r}}^2 \Big[ \tilde{h}''_{c} + \frac{1}{\tilde{\tilde{r}}} \tilde{h}'_{c} - \frac{1}{\tilde{\tilde{r}}^2} \tilde{h}_{c} \Big] + k_z^2 \Big( 2 + \tilde{\tilde{r}}^2 \Big) \tilde{h}_{c} = & 2 k_z \tilde{\tilde{r}} h_{_{LL}}. \label{eqn:magcorrDrzero}
\end{eqnarray}

\begin{figure}
\begin{center}
\includegraphics[scale=0.125]{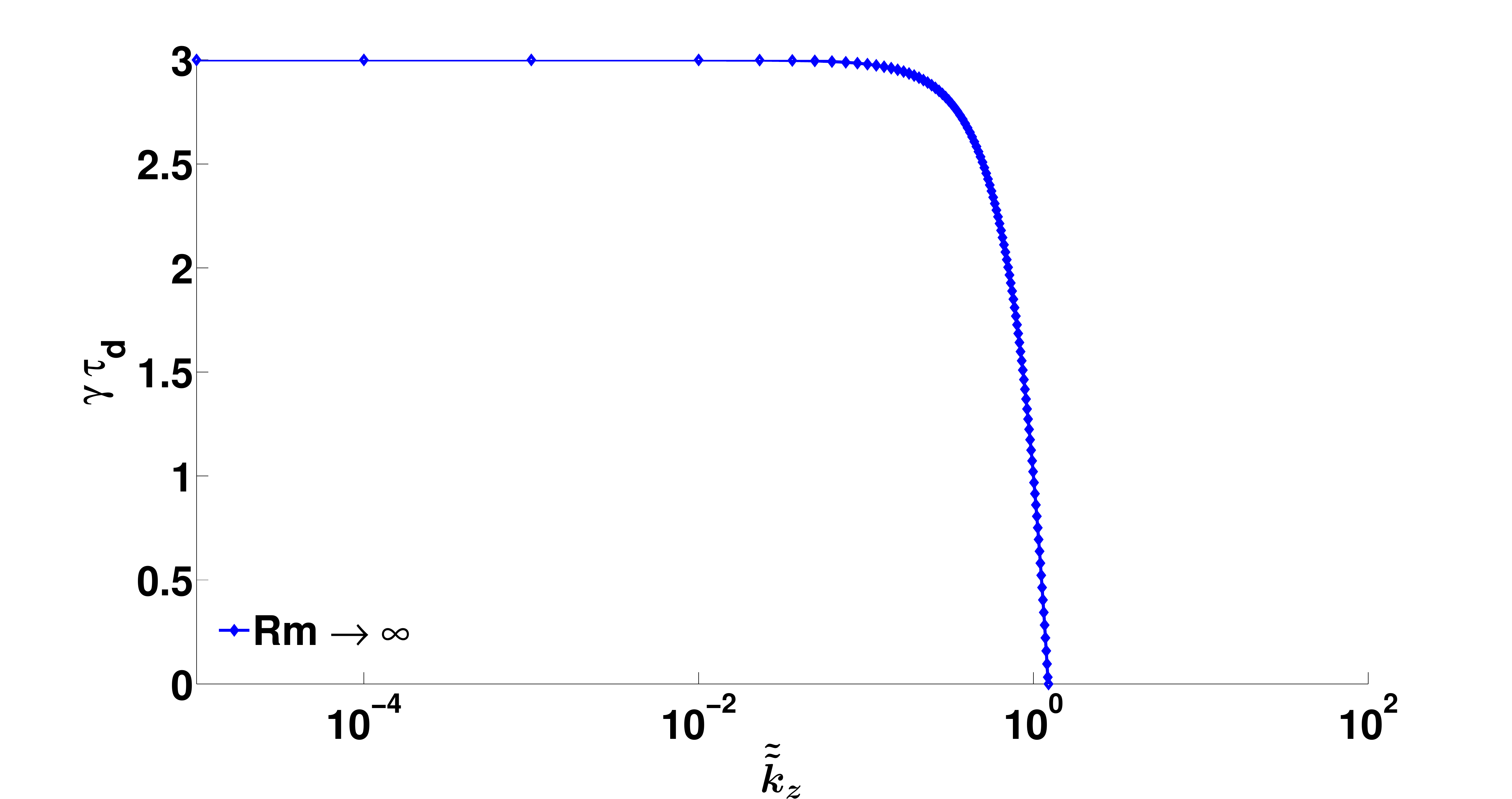}
\includegraphics[scale=0.125]{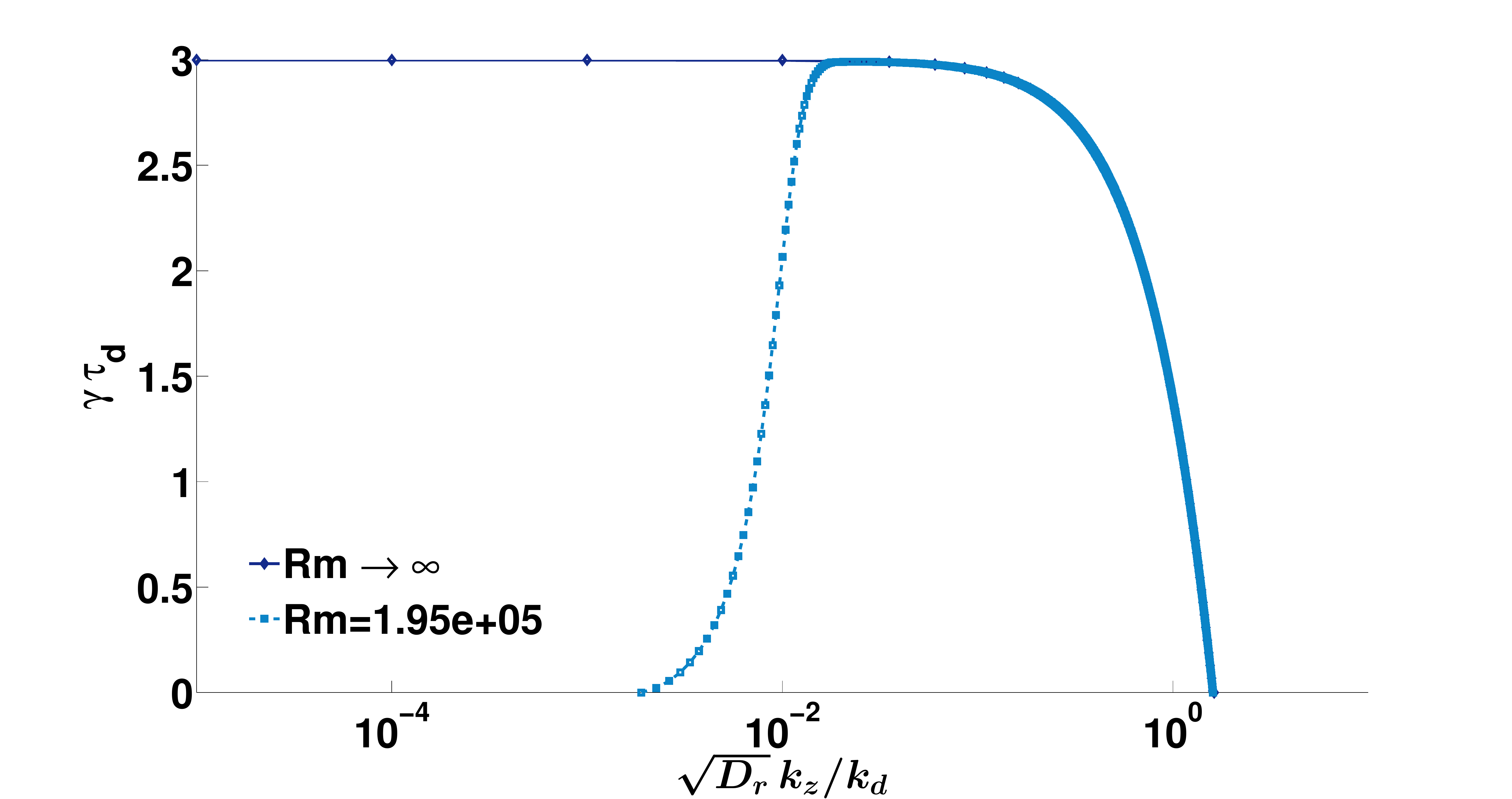}
\end{center}
\caption{Growth rate as a function of the rescaled $k_z$ for the case of 1. on the left panel the limit $D_r \rightarrow 0$ where only the infinite $Rm$ has a dynamo effect, 2. on the right the limit $D_r \rightarrow \infty$ where both the finite and the infinite $Rm$ have dynamo instability.}
\label{fig:LargeSmallDr}
\end{figure}

The eigenvalues of these equations gives the asymptotic behaviour of the growth rate when first the limit $Rm \rightarrow \infty $ is taken and then the limit $D_r \rightarrow 0$. 
The resulting eigenvalues from the above set of equations are shown separately in the left panel of figure \ref{fig:LargeSmallDr}. 
These results are valid provided that $1 \gg D_r \gg Rm^{-1}$, but the expansion fails if $D_r$ is the same order as $Rm^{-1}$.
For values of $D_r$ smaller than this threshold the dissipation effects are stronger and the dynamo instability disappears. 

\subsection{$Rm\rightarrow \infty, D_r \rightarrow \infty$}

 In the top panel of figure \ref{fig:becom2d} as the parameter $D_r \rightarrow \infty$ we see that the unstable $k_z$ modes move towards smaller values. 
This implies that the magnetic field should be correlated over longer distances along the $z$ direction in order for a large $u_z$ to twist and fold the 
field lines and result in the amplification of the magnetic field. A similar behaviour is observed in the case of infinite $Rm$ ($Rm \rightarrow \infty$), 
shown in the bottom panel of figure \ref{fig:becom2d}. It is important to note here that the growth rate $\gamma$ is non-dimensionalized with $D_1$ which is related 
to the amplitude of the shear in the correlation function $g_{_{2D}}$. If the growth rate is normalized with $\sqrt{D_1^2+D_2^2}$ which takes into account 
both the shear in in $u_{_{2D}}$ and $u_z$ then the normalized growth rate
$\gamma/\sqrt{D_1^2+D_2^2}=\gamma\tau_d/\sqrt{1+D_r^2}$ becomes zero in the limit $D_r\to\infty$. 
Thus there is no violation of the anti-dynamo theorem. The maximum growth rate in figure \ref{fig:becom2d} appears to be independent of $D_r$ in the large $D_r$ limit. 
The growth rate curves for large $D_r$ can be plotted with a rescaled $k_z \rightarrow \sqrt{D_r}k_z$ which make the curves to collapse on each other (not shown here). 
Such a result can be obtained by expanding the equations \ref{eqn:magcorreqns} in terms of $1/D_r$ and solving for the lowest order equations which represents 
the limit $D_r \rightarrow \infty$. Since the steps are similar with the previous section the resulting set of equations are not shown. 
 
 The eigenvalues of the resulting equations after taking the limit $D_r \rightarrow \infty$ are shown in the right panel of figure \ref{fig:LargeSmallDr}. In this plot we show both the finite $Rm$ and the infinite $Rm$ growth rates. The behaviour of the two curves are similar except for the small $k_z$ where the finite $Rm$ limit looses the dynamo instability as shown in section \ref{Subsec:kzzero}.
For fixed $k_z/k_d$, however, the limits $\lim_{Rm \rightarrow \infty}$ and $\lim_{D_r \rightarrow \infty}$ are commuting:
\begin{align}
\lim_{Rm \rightarrow \infty} \lim_{D_r \rightarrow \infty} \gamma = \lim_{D_r \rightarrow \infty} \lim_{Rm \rightarrow \infty} \gamma.
\end{align}

\section{Correlation functions and energy spectra} \label{Section:Six}   

In this section we discuss the functional form of the correlation functions and the spectra of the most unstable eigenmode. 
It is reminded that the magnetic energy spectra of a magnetic field advected by a Kazantsev $2D$ flow 
show the power-law behaviour $k^2$ for wavenumbers between the velocity wavenumber $k_0$ and the dissipation wavenumber $k_d$.
%
While in $3D$ the spectrum of the unstable mode scales like $k^{3/2}$ in the same range. 
For the $2.5D$ problem there are $3$ relevant scales $k_z, k_d, k_0$. Dynamo instability is obtained only for a particular ordering of these scales. 
Based on the results from the previous sections, to obtain a dynamo $\sqrt{D_r} k_z$ cannot be much larger than $k_d$ nor much smaller than $k_0$, more precisely  
$c_{min} k_0 \leq \sqrt{D_r} k_z < c_{max} k_d$. 
The two constants $c_{min}$ and $c_{max}$ are related to $k_{min}$ and $k_{max}$ respectively discussed in section \ref{Section:Four}.
It is found that $c_{min}$ depends on the $Rm$ and $c_{max} \approx 1.6$ calculated for large $Rm$. 
We concentrate on the case of $Rm \rightarrow \infty$ where we have two scales in the system $k_d, k_z$. 
First we examine the behaviour of the correlation functions $h_{_{LL}} \left( r \right), h_c \left( r \right)$ before moving to the spectra of the magnetic field.

We start with equations \ref{eqn:magcorrinf}, for $Rm \rightarrow \infty$ where the equations are written in terms of the rescaled quantities $\tilde{r}, \tilde{k_z}$. The dissipation scale $r_d = 1/k_d$ is given by $\tilde{r} = 1$. The small and large $\tilde{r}$ asymptotics of $h_{_{LL}} \left( \tilde{r} \right), h_c \left( \tilde{r} \right)$ are mentioned in Appendix \ref{Appendix:Two}. There are three distinct range of scales that display different behaviour. The small $\tilde{r}$ corresponds to the regime of scales below the dissipation scale $\tilde{r} \ll 1$, the large $\tilde{r}$ corresponds to the regime ${\tilde{r} \gg 1/\tilde{k}_z}$. In between these two range of scales we have an intermediate range of scales $1 \ll \tilde{r} \ll 1/k_z$. 
The scaling in this range of scales can be obtained by using matched asymptotics, the details of which are given in the Appendix \ref{Appendix:Four}. In this process we also find that in the limit of $\tilde{k}_z \rightarrow 0$ we can obtain the eigenvalue $\gamma \rightarrow 3$ independent of the value of $D_r$, in accordance with results shown in figures \ref{fig:Growth1}, \ref{fig:becom2d}. The correlation functions $h_{_{LL}}\left( r \right), {h}_c \left( r \right)$ show the following scaling with the variable $r$ for the large $Rm$ limit,
\begin{align}
h_{_{LL}} = \begin{cases} 
   1 - c_1 r^{2} + O(r^4) & \text{if } r \ll \frac{1}{k_d} \\
   c_2 r^{-1} & \text{if } \frac{1}{k_d} \ll r \ll \frac{1}{k_z} \\
   e^{-c_3 r} & \text{if } r \gg \frac{1}{k_z}
  \end{cases}, \hspace{5mm}
{h}_{c} = \begin{cases} 
   c_4 r^{1} & \text{if } r \ll \frac{1}{k_d} \\
   c_5 r^{0} & \text{if } \frac{1}{k_d} \ll r \ll \frac{1}{k_z} \\
   e^{-c_2 r} & \text{if } r \gg \frac{1}{k_z}
  \end{cases}\label{eqn:behavCorr}
\end{align} 
where $c_1, c_2, c_3, c_4, c_5$ are related to $\eta, k_z, D_r$ and can be found from the calculation in Appendix \ref{Appendix:Four}. In figure \ref{fig:structfuncts} we show the correlation functions $h_{_{LL}}(\tilde{r}), {h}_c(\tilde{r})$ for $\tilde{k}_z = 0.005, D_r = 1$. Since the equations are rescaled with $k_d$ the dissipation scale is given by $\tilde{r} = 1$. We can see that the behaviour of the functions $h_{_{LL}}\left( \tilde{r} \right), {h}_c \left( \tilde{r} \right)$ described in equation \ref{eqn:behavCorr} is well captured from the numerics. 

\begin{figure}
\begin{center}
\includegraphics[scale=0.2]{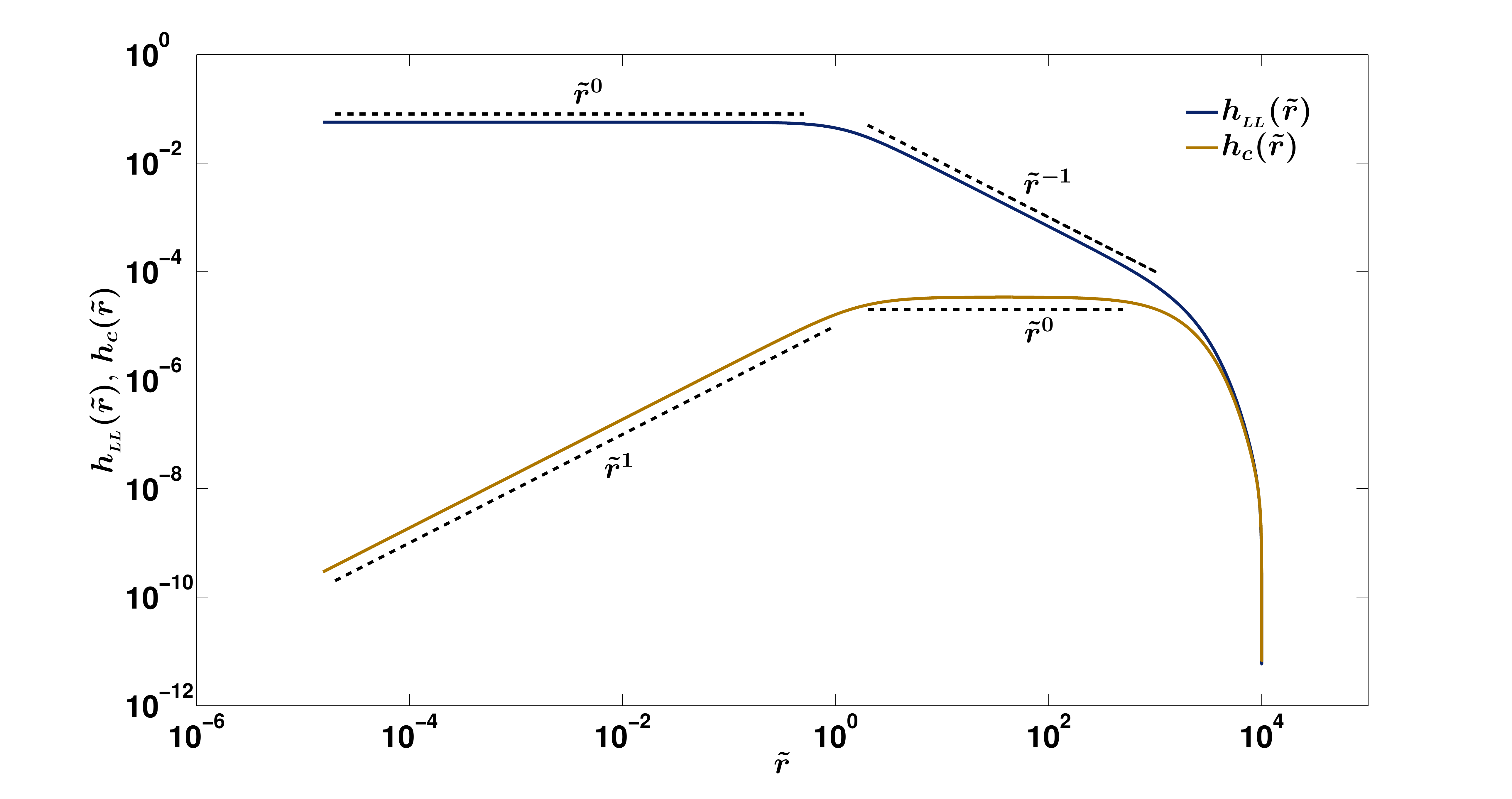}
\end{center}
\caption{The figure shows the correlation functions of the magnetic field $h_{_{LL}}(\tilde{r})$ in dark blue, ${h}_c(\tilde{r})$ in light brown for $k_z = 0.005$ with $D_r = 1$. The black dashed lines denote exponents that are observed in the respective range of scales.}
\label{fig:structfuncts}
\end{figure}

\begin{figure}
\begin{center}
\includegraphics[scale=0.3]{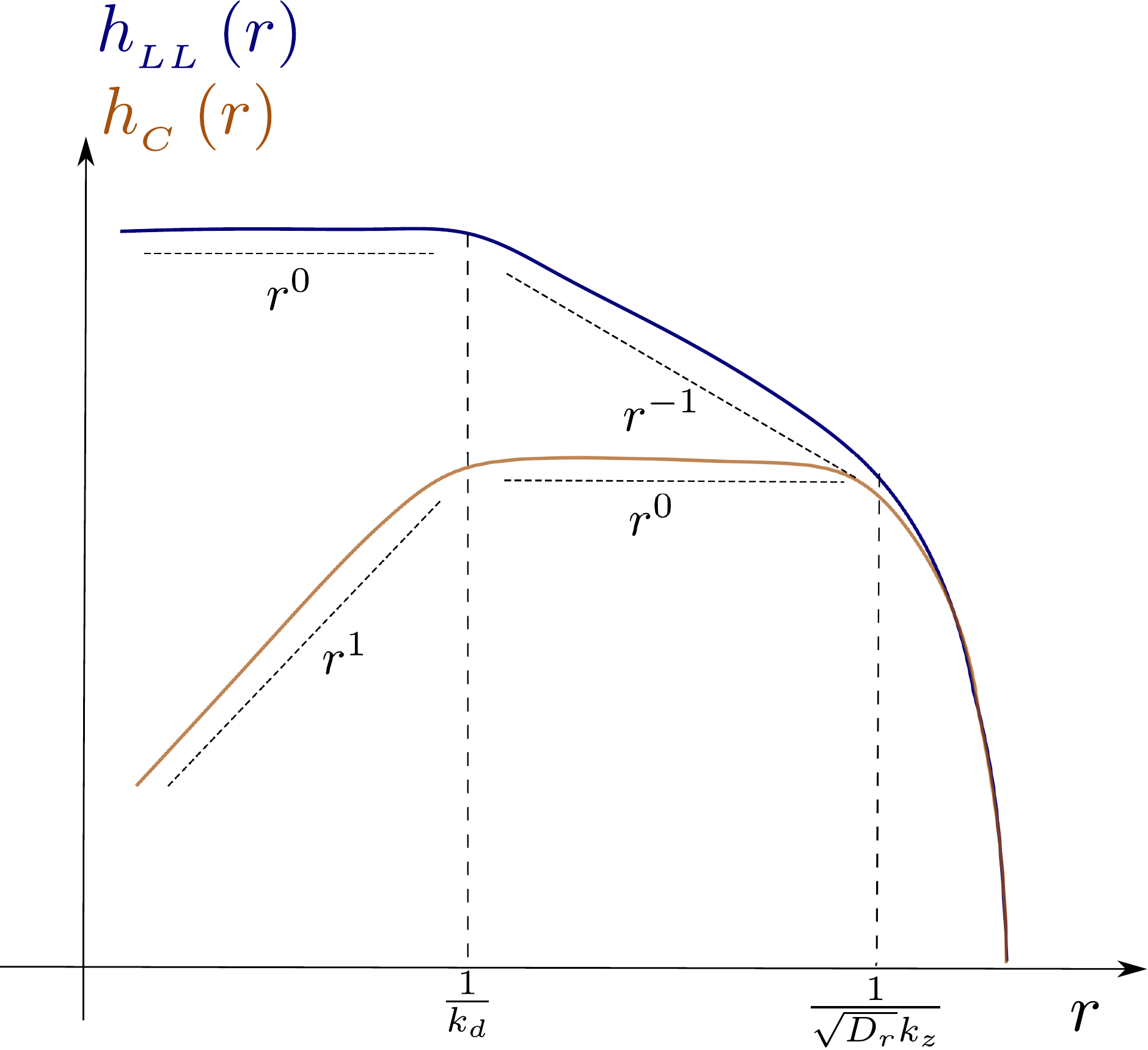}
\includegraphics[scale=0.3]{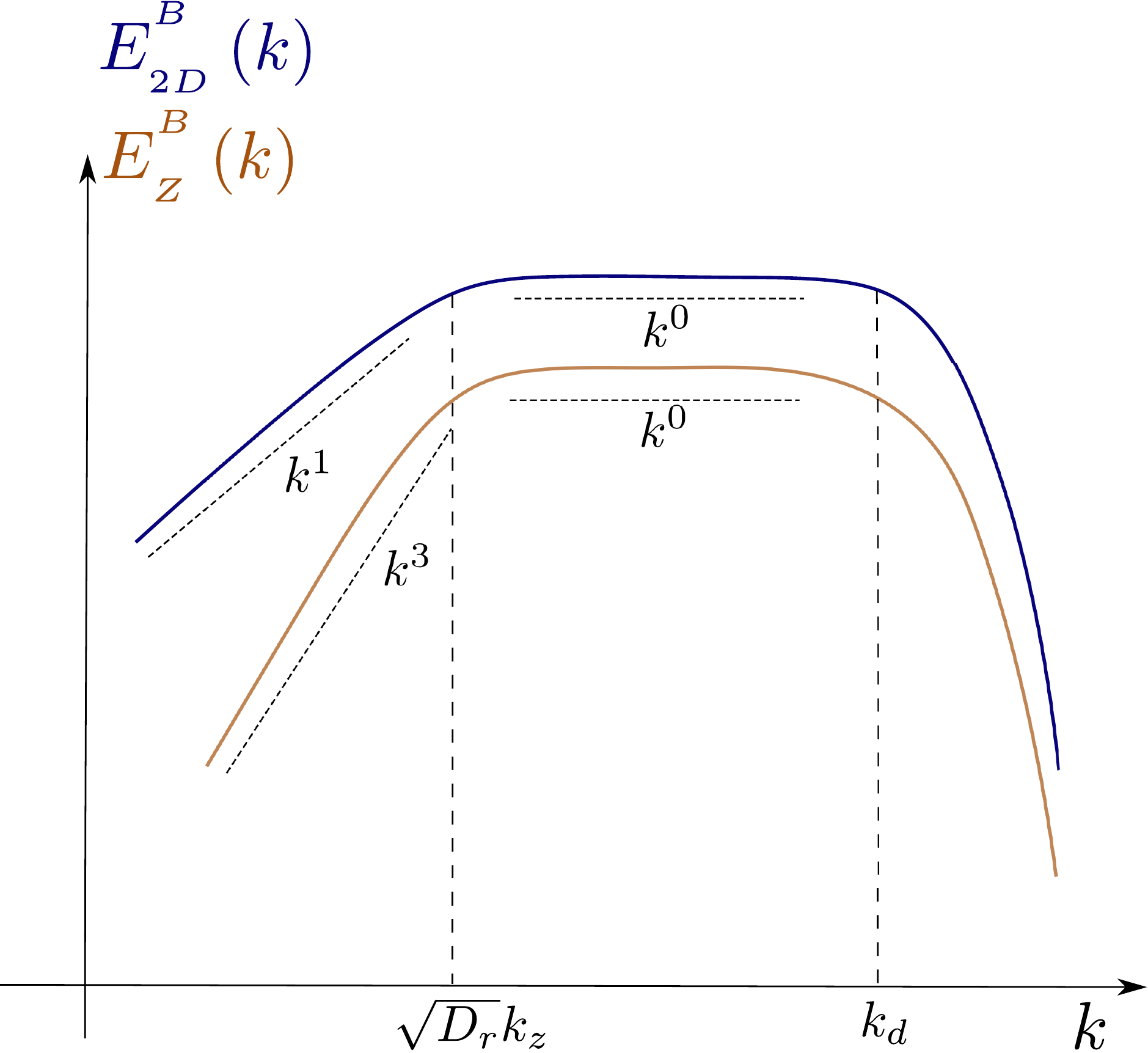}
\end{center}
\caption{The figures above shows the form of 1. on the left of the correlation functions $h_{_{LL}} \left( r \right)$ in dark shade blue , $h_c \left( r \right)$ in light shade brown and 2. on the right the spectra $E^{B}_{_{2D}} \left( k \right)$ in dark shade blue, $E^{B}_{_{Z}} \left( k \right)$ in light shade brown. The black dashed lines represents the different exponents which are observed in the respective range of scales.}
\label{fig:Plotsmade}
\end{figure}

Now with the solution of $h_{_{LL}}(r), {h}_c(r)$ we can construct the spectra using the Wiener-Khinctine relation (see \cite{chatfield1989analysis}) in two dimensions. For a function $M(r)$ its isotopic Fourier spectrum reads as, 
\begin{align}
\widehat{M} \left( k \right) = k \int_{0}^{\infty} r M(r) J_0 \left( k r \right) dr.
\end{align}
For the magnetic field we can construct the planar magnetic field spectrum $E^{B}_{_{2D}} \left( k \right)$ and the vertical magnetic energy spectrum $E^{B}_{_{Z}} \left( k \right)$. Their relations with $h_{_{LL}}(r), {h}_c(r)$ are given by,
\begin{align}
E^{B}_{_{2D}} \left( k \right) & = k \int_{0}^{\infty} r \Big( 2 h_{_{LL}}(r) + r h_{_{LL}}'(r) + r k_z {h}_c(r) \Big) J_0 \left( k r \right) dr \\
E^{B}_{_{Z}} \left( k \right) & = k \int_{0}^{\infty} r \frac{1}{k_z} \Big( {h}_c'(r) + \frac{{h}_c(r)}{r} \Big) J_0 \left( k r \right) dr
\end{align}
Using the behaviour of the correlation functions $h_{_{LL}}(r), {h}_c(r)$ mentioned in equation \ref{eqn:behavCorr} we can use the Wiener-Khintchine relation to get the behaviour of $E^{B}_{_{2D}} \left( k \right), E^{B}_{_{Z}}\left( k \right)$ in the regimes $k \ll k_z, k_z \ll k \ll k_d, k \gg k_d$. We can write the generalized spectra of the magnetic field in the limit of large scale separation $k_z \ll k_d$ as,
\begin{align}
E^{B}_{_{2D}} \left( k \right) = \begin{cases} 
   k^{1} & \text{if } k \ll k_z \\
   k^{0} & \text{if } k_z \ll k \ll k_d \\
   e^{-k/k_d} & \text{if } k \gg k_d
  \end{cases}, \hspace{5mm}
E^{B}_{_{Z}} \left( k \right) = \begin{cases} 
   k^{3} & \text{if } k \ll k_z \\
   k^{0} & \text{if } k_z \ll k \ll k_d \\
   e^{-k/k_d} & \text{if } k \gg k_d
  \end{cases}
\end{align}
\begin{figure}
\begin{center}
\includegraphics[scale=0.125]{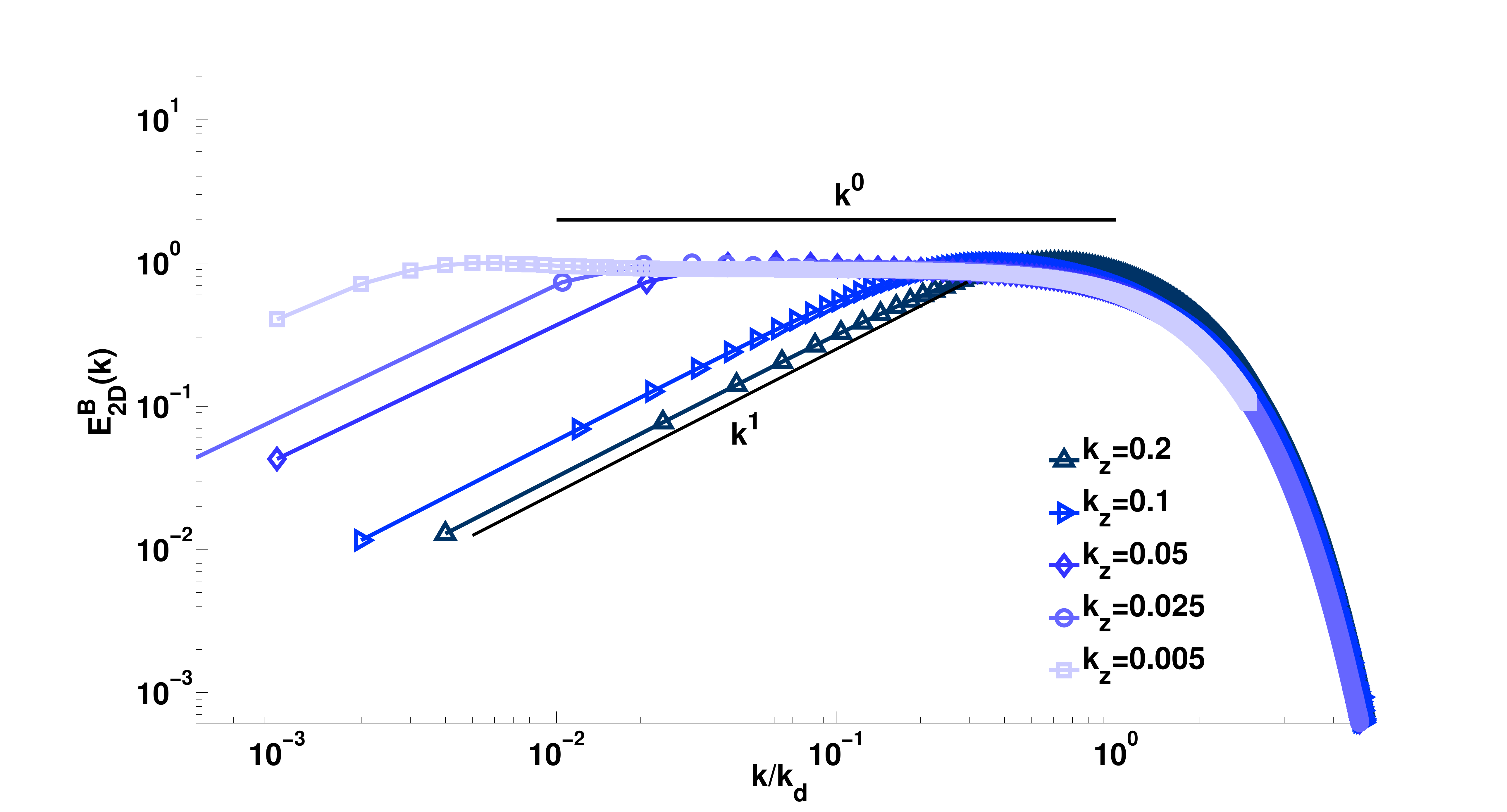}
\includegraphics[scale=0.125]{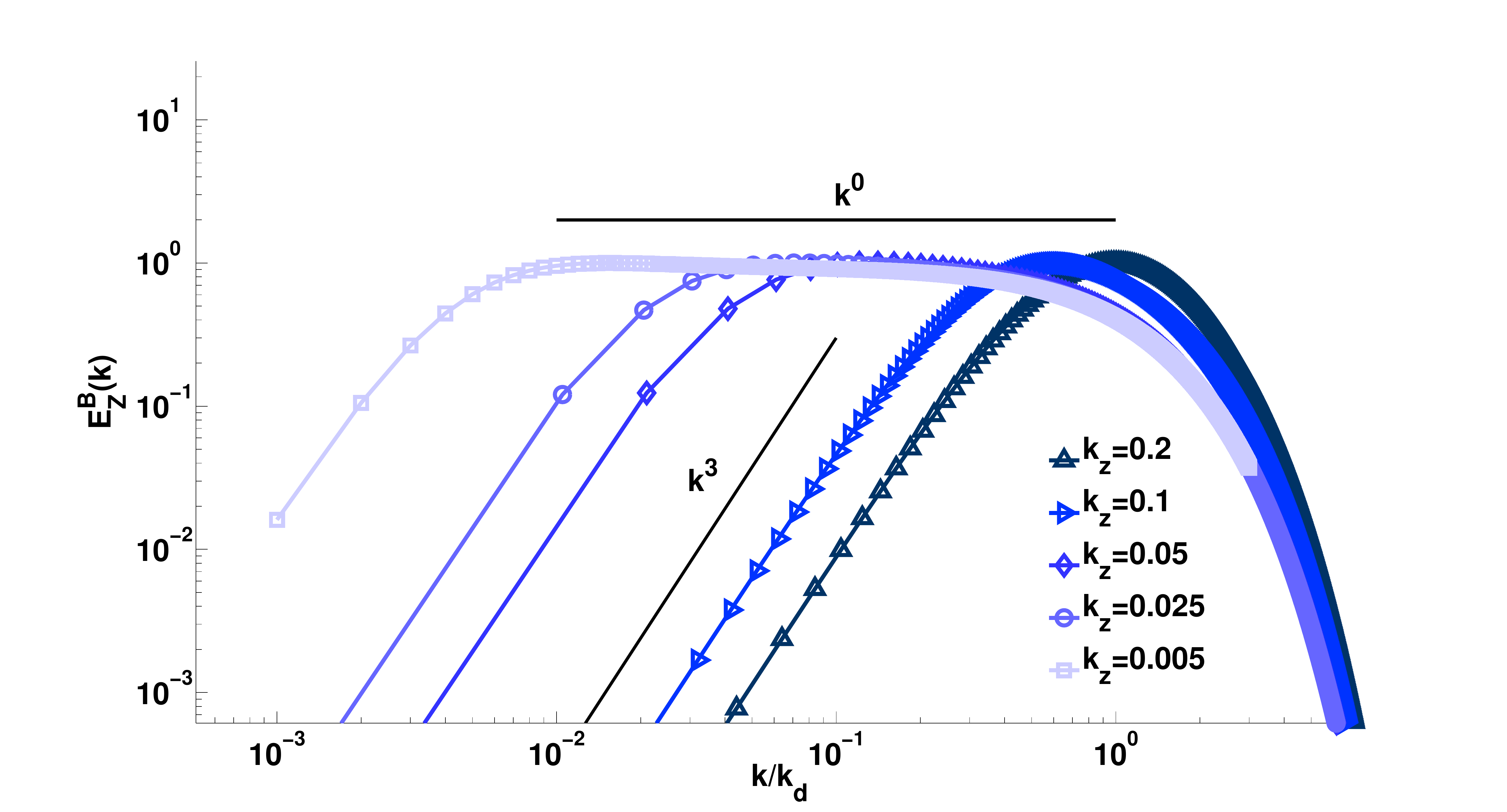}
\end{center}
\caption{Spectra of the magnetic field 1. on the left - $E^{B}_{_{2D}} \left( k \right)$, 2. on the right $E^{B}_{_{Z}} \left( k \right)$ for different values of $k_z$ shown in the legends. Lighter shades of blue correspond to smaller values of $k_z$. The parameter $D_r = 1$, the black lines denote power laws.}
\label{fig:Spectras}
\end{figure}
These predicted power laws are in agreement with the solutions of the equations \ref{eqn:magcorrinf} displayed in figure \ref{fig:Spectras}. 
In this figure the dissipation wavenumber is unity and $k_z$ is varied with the values mentioned in the legend.
Figure \ref{fig:Plotsmade} summarizes the form of the unstable mode for the different range of scales in both $k$ and $r$ for the case of large scale separation $k_z \ll k_d$ and generalized to take into account the variation in $D_r$. 

\section{Comparison with direct numerical simulations} \label{Section:Seven}

\subsection{White noise flows}

In order to test the relevance of the theoretical results with the results of
Direct Numerical Simulations (DNS) we consider and solve numerically the partial differential equation \ref{eqn:induction} for 
a random Gaussian distributed flow in a finite two dimensional periodic box. 
We note that the two dimensional periodic flow does not respect isotropy. 
This is true for any finite homogeneous system, 
thus we will be limited to only a qualitative comparison. We consider a random flow of the form, 
\begin{align}
\psi \left( x, y, t \right) & = \zeta_3 \left( t \right) \Big[ \sin \left( \phi_3 \left( t \right) \right) \cos \left( k_f \, x + \phi_4 \left( t \right) \right) + \cos \left( \phi_3 \left( t \right) \right) \sin \left( k_f \, y + \phi_4 \left( t \right) \right) \Big]/k_f \label{eqn:simvel1} \\
u_z  \left( x, y, t \right) & = \zeta_4 \left( t \right) \Big[ \sin \left( \phi_3 \left( t \right) \right) \sin \left( k_f \, x + \phi_4 \left( t \right) \right) + \cos \left( \phi_3 \left( t \right) \right) \cos \left( k_f \, y + \phi_4 \left( t \right) \right) \Big] \label{eqn:simvel2}
\end{align}
where $\zeta_3 \left( t \right), \zeta_4 \left( t \right)$ are two Gaussian distributed random variables satisfying the relations, $\left\langle \zeta_3 \left( t \right) \zeta_3 \left( t' \right) \right\rangle = \delta \left( t - t' \right)$, $\left\langle \zeta_4 \left( t \right) \zeta_4 \left( t' \right) \right\rangle = \delta \left( t - t' \right)$, $\left\langle \zeta_3 \left( t \right) \zeta_4 \left( t' \right) \right\rangle = 0$. $\phi_3 \left( t \right), \phi_4 \left( t \right)$ are uniformly distributed random variables in the interval $[0, 2 \pi]$. The above flow is realized in a domain $[2 \pi L, 2 \pi L]$ with $k_f\, L$ being the forcing wavenumber. The above system is homogeneous and invariant under $\pi/2$ rotations. The discretized version of the induction equation is numerically solved with the realization of the noise changing at each time step with the Stratonovich formulation of the noise (see \citep{greiner1988numerical, leprovost2004influence}).

\begin{figure}
\begin{center}
\includegraphics[scale=0.11]{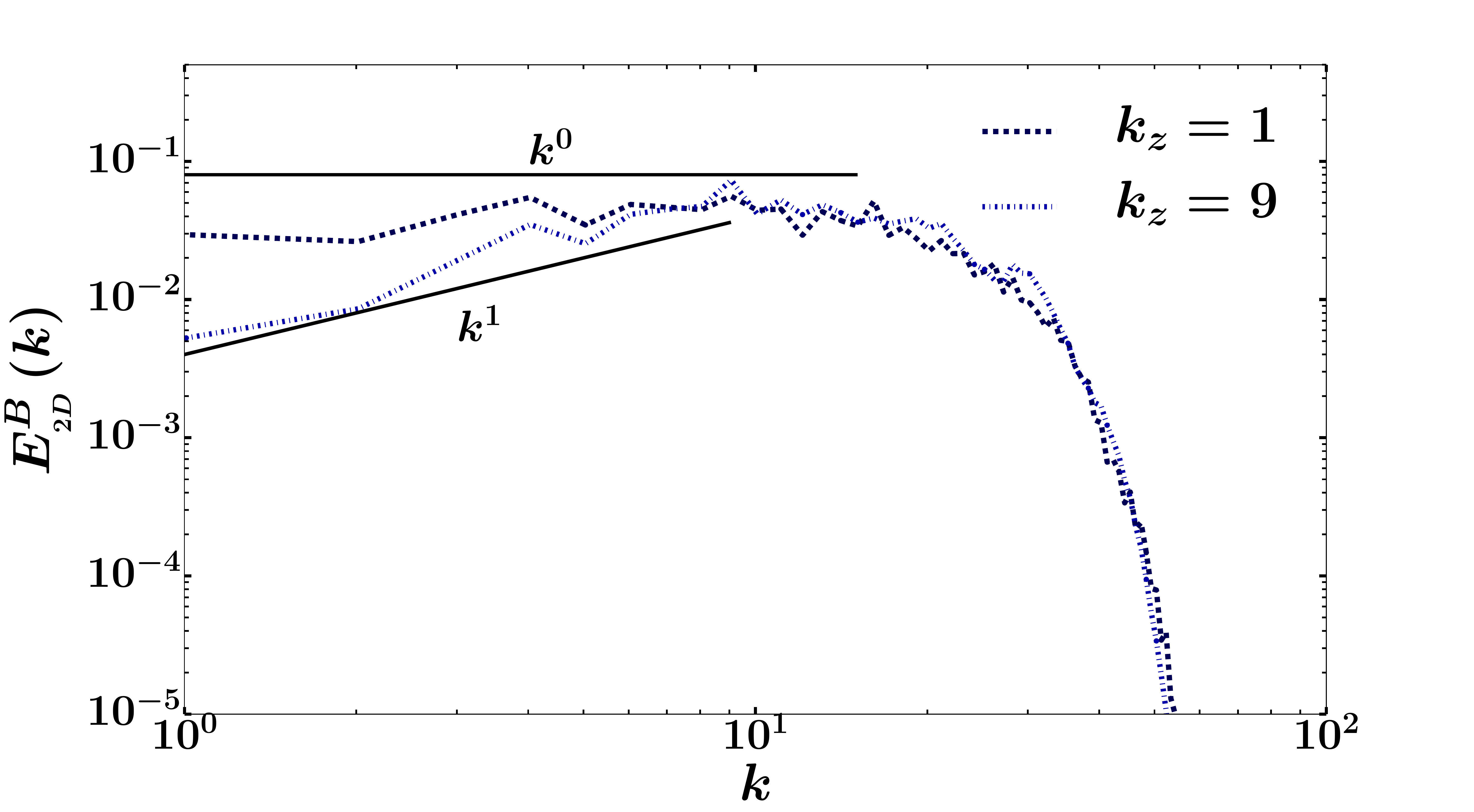}
\includegraphics[scale=0.11]{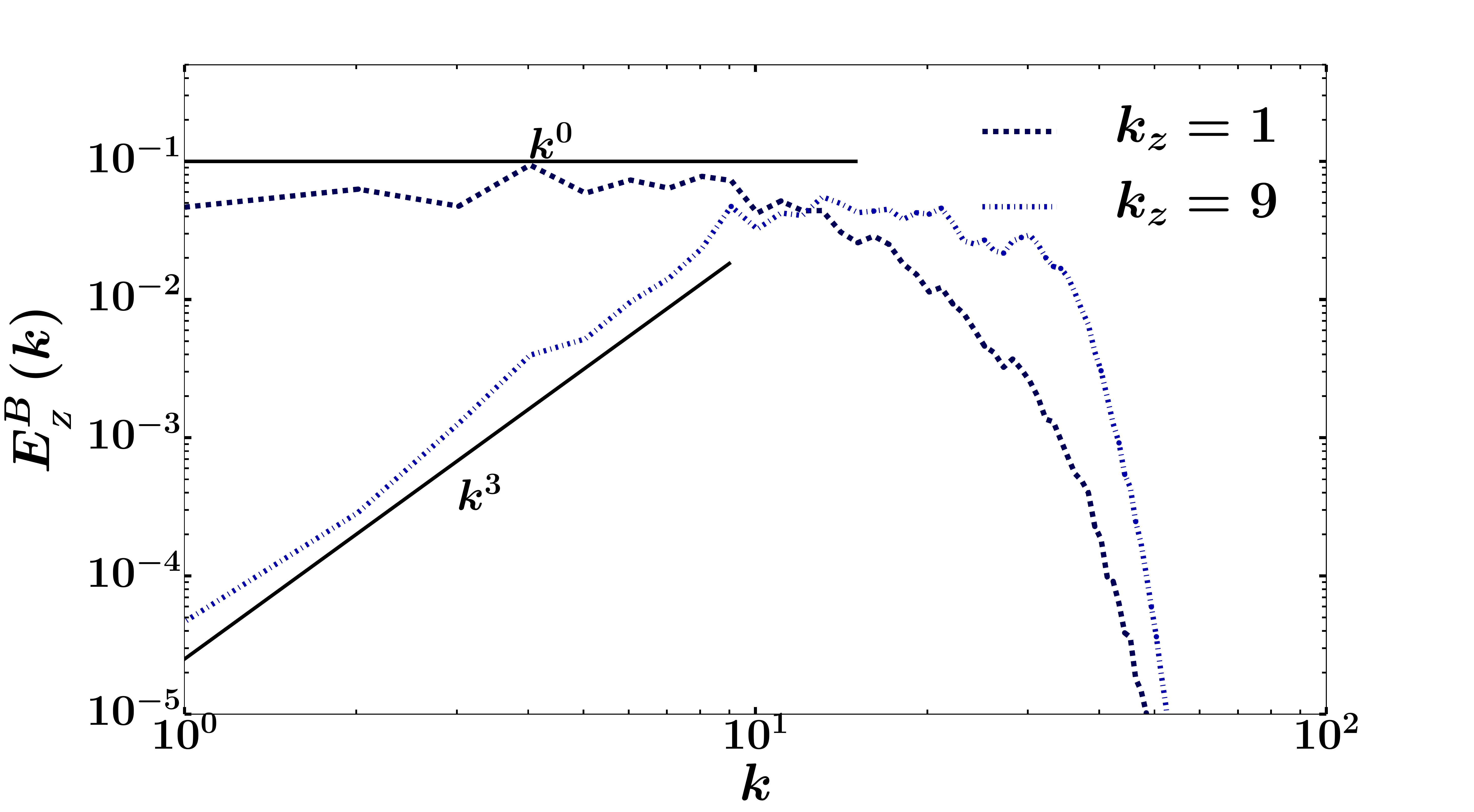}
\end{center}
\caption{The figures above show the magnetic field spectrum at one instant of time 1.) $E_{_{2D}}^{B}$ on the left, 2.) $E_{_{Z}}^B$ on the right for a few different values of $k_z$ mentioned in the legend, lighter shades correspond to increasing values of $k_z$. The results correspond to the fluctuating velocity field with parameters $Rm \approx 460$. }
\label{fig:spectrasimulfluc}
\end{figure}

\begin{figure}
\begin{center}
\includegraphics[scale=0.3]{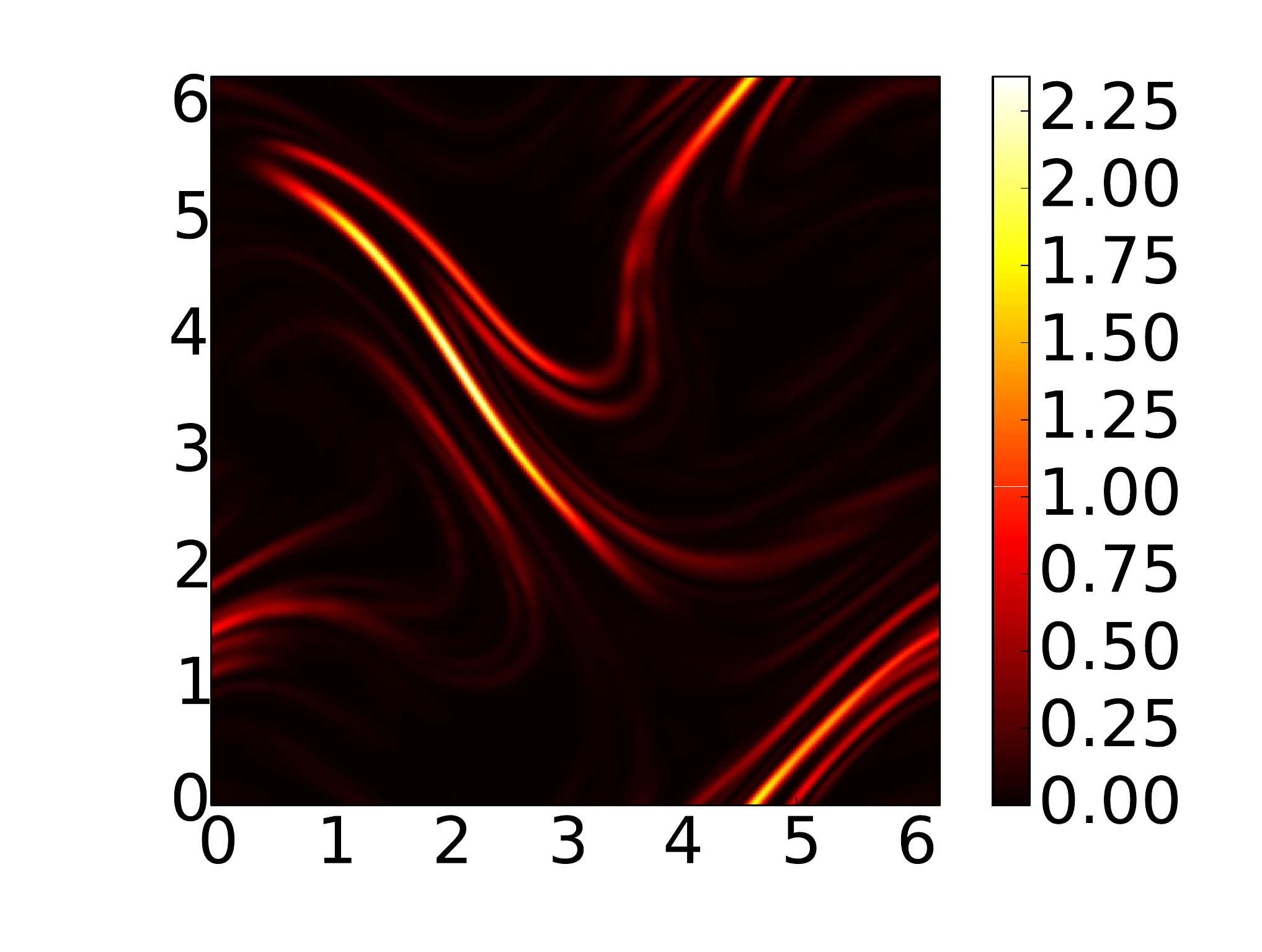}
\includegraphics[scale=0.3]{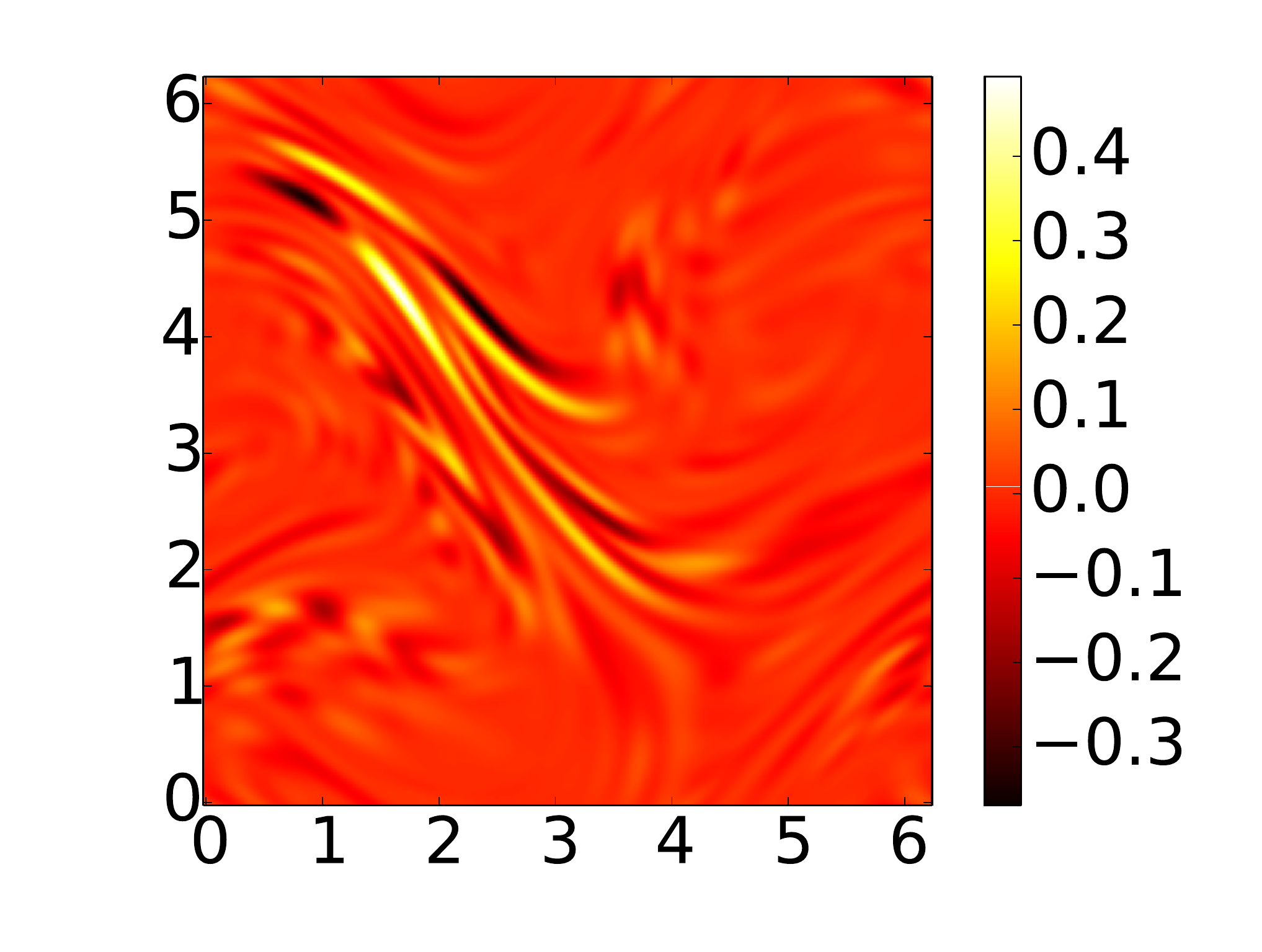}
\end{center}
\caption{The figures above show the contour of the magnetic field  with 1.) $|{\bf b}_{_{2D}}|^2$ on the left, 2.) the real part of $b_z$ on the right. The results correspond to the fluctuating velocity field with parameters $Rm \approx 210$ and $k_z/k_d \approx 0.35$. }
\label{fig:vismagfield}
\end{figure}

The growth rate calculated for the magnetic field with $k_f = 1$ is shown in figure \ref{fig:Timecompare} on the left panel for a few values of $Rm$. Qualitatively the results reproduce the behaviour of the theoretical predictions. The spectra of the growing magnetic field are shown in figure \ref{fig:spectrasimulfluc} for a single time realizations for a $Rm \approx 460$ and a few values of $k_z$ mentioned in the legend. The theoretical predictions are shown in black solid lines and they compare well with the numerical results. The magnetic field intensity is shown in figure \ref{fig:vismagfield}, where the panel on the left shows $|{\bf b}_{_{2D}}|^2 = b_x^{\dagger} b_x + b_y^{\dagger} b_y$ the magnetic energy in the $2D$ plane and the panel on the right shows the real part of the vertical magnetic field $b_z$. The magnetic field lines are concentrated in thin filamentary structures and their size decreases as $Rm$ is increased.

\subsection{Freely evolving flows}

To test the validity of the model for more realistic flows 
we also compare our results with the growth rates of freely evolving chaotic/turbulent flows. 
We consider a flow driven by a non-helical forcing at a wavenumber $k_f=4$ that is constant in time.
The temporal behaviour of the flow and its `randomness' originates purely from the chaotic dynamics of the Navier-Stokes equation.
The details of the full study of this system of equations can be found in \cite{seshasayanan2015turbulent}. 

\begin{figure}
\begin{center}
\begin{tabular}{p{7cm} p{7cm}}
\subfloat{\includegraphics[scale=0.13]{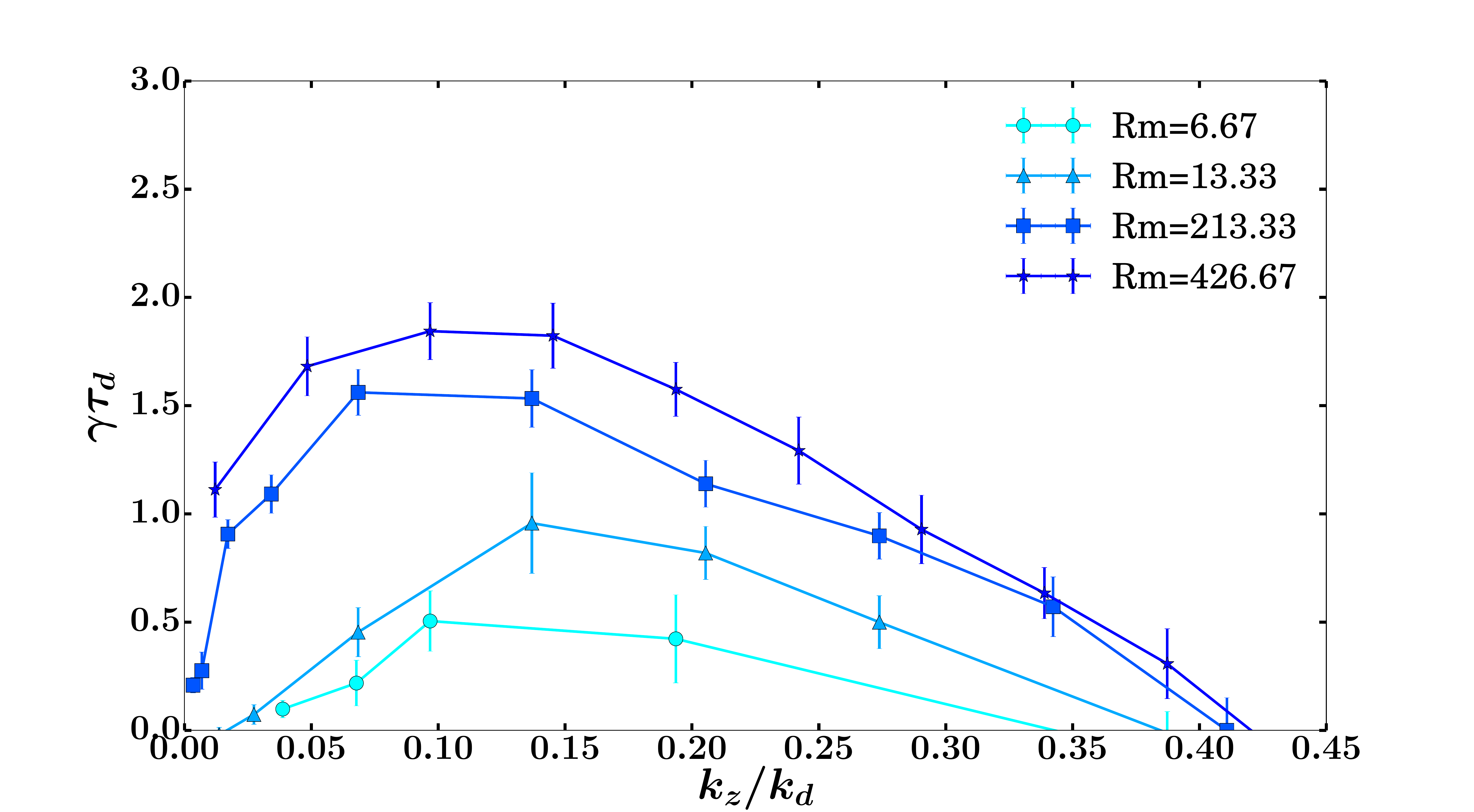}}
&\subfloat{\includegraphics[width=6.2cm, height=4cm]{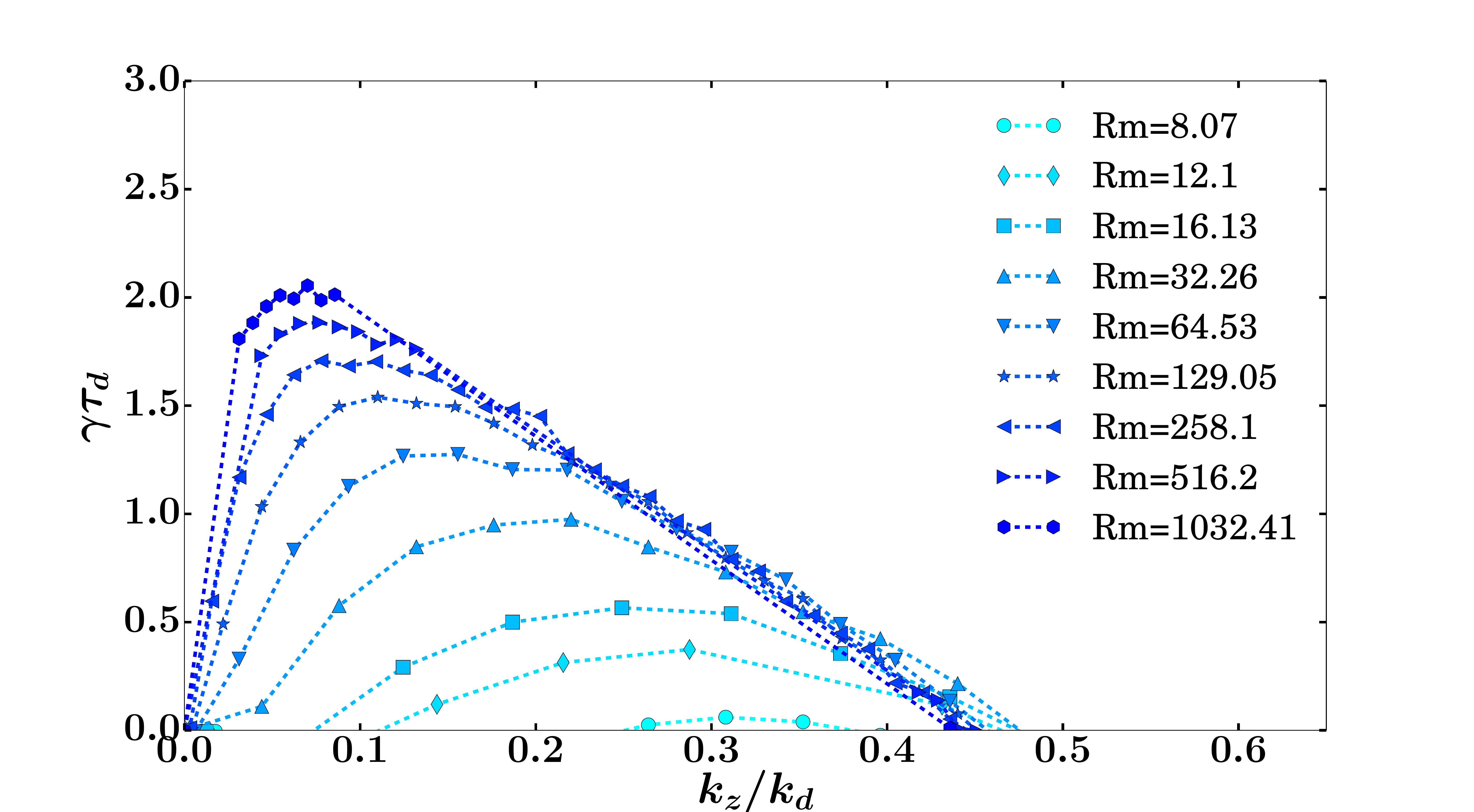}}
\end{tabular}
\end{center}
\caption{The figures above show the growth rate $\gamma$ as a function of the normalized wavemode $k_z/k_d$ for 1. on the left for the delta correlated flow and 2. on the right for a time correlated flow. Darker shades correspond to larger values of $Rm$.}
\label{fig:Timecompare}
\end{figure}

\begin{figure}
\begin{center}
\includegraphics[scale=0.11]{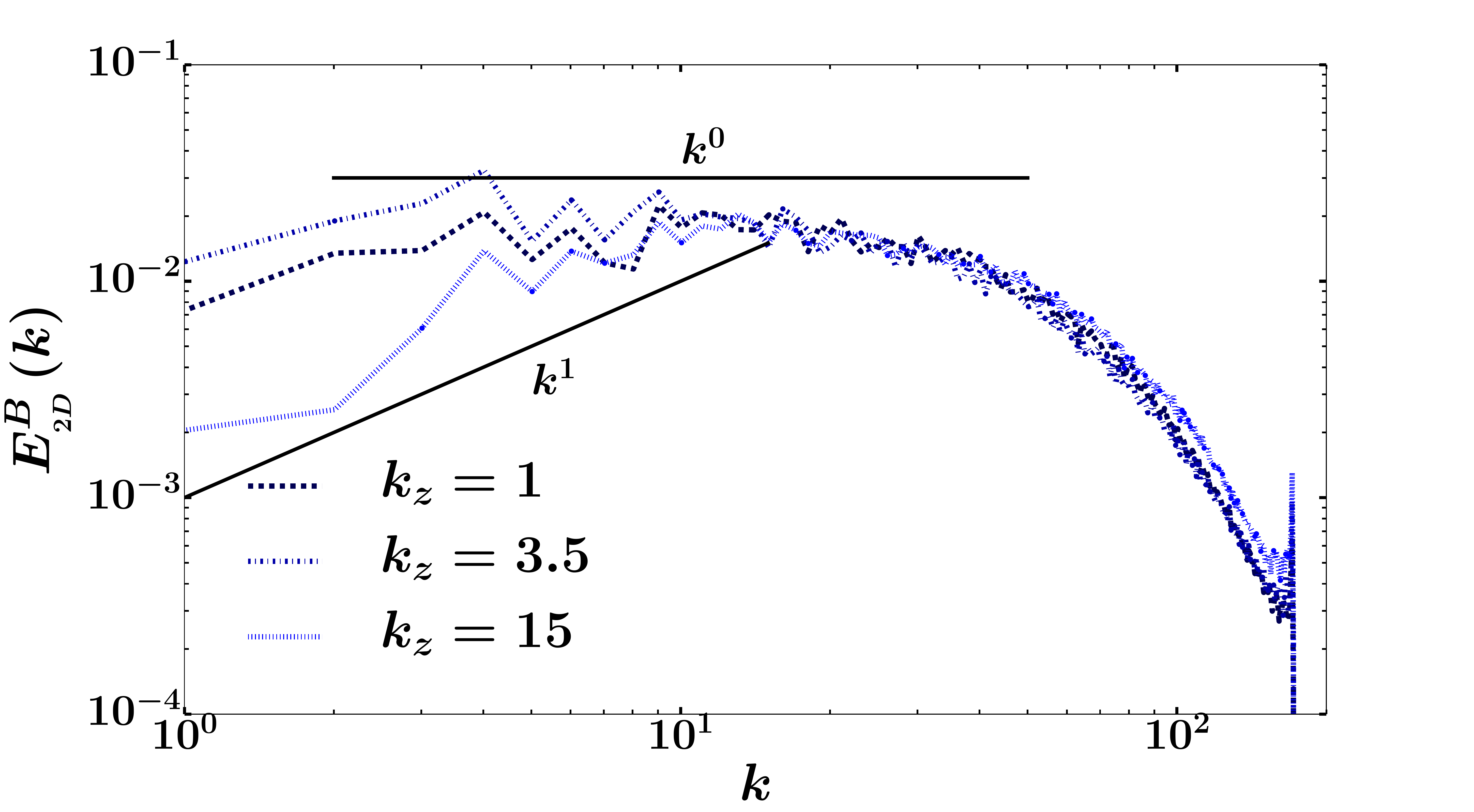}
\includegraphics[scale=0.11]{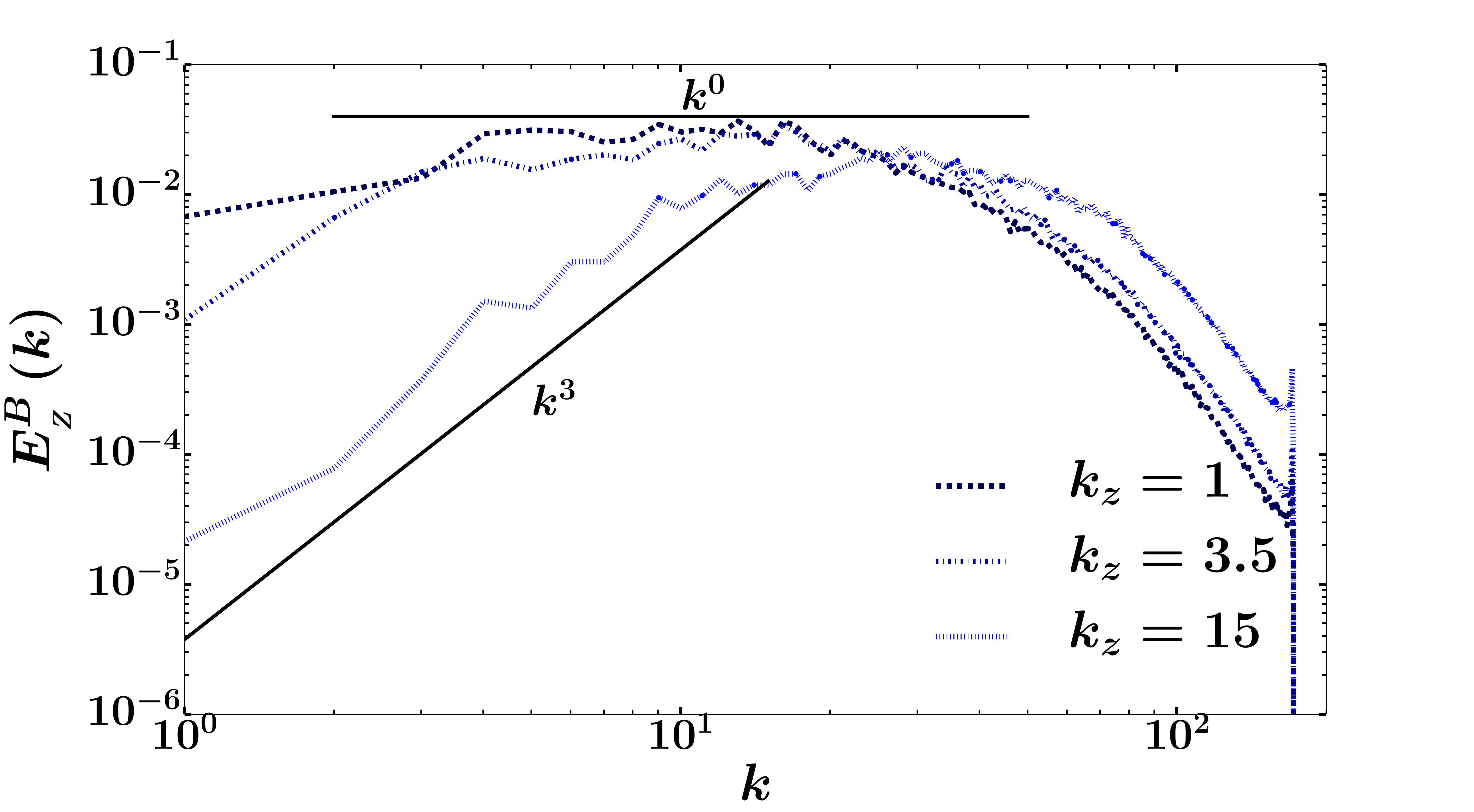}
\end{center}
\caption{The figures above show a time shot of the magnetic field spectra 1.) $E_{_{2D}}^{B}$ on the left, 2.) $E_{_{Z}}^B$ on the right for a few different value of $k_z$ mentioned in the legend, lighter shades correspond to increasing values of $k_z$. The results correspond to the kinematic dynamo problem of the forced Navier Stokes equation with parameters $Rm \approx 1020, Re \approx 32$. }
\label{fig:spectrasimulforc}
\end{figure}

The normalized growth rate $\gamma$ obtained from the turbulent flow is shown in the right panel in Figure \ref{fig:Timecompare}
as a function of the normalized $k_z/k_d$ and for different values of $Rm$. 
For the examined flow the quantity $k_d = u/\eta = k_f \sqrt{Rm}$ and $Rm = u/(k_f \eta), \tau_d = 1/(\eta k_d^2)$ where $u$ is the r.m.s velocity. 
We find a good match in terms of the behaviour of the growth rates and its dependence on $k_z/k_d, Rm$. 
The spectra of the magnetic field, $E^{B}_{2D}$ and $E_{_{Z}}^{B}$ are also shown in figure \ref{fig:spectrasimulforc} along with the black solid lines 
denoting theoretical prediction mentioned in the previous section. The spectra shown correspond to a simulation run with the parameters 
$Rm \approx 1020, Re \approx 32$ taken after $t \approx 100$ non-linear time scale. The theoretical predictions seem to capture well the shape of the unstable spectra. 

\section{Conclusions} \label{Section:Eight}

In this work we have examined the dynamo properties of the Kazantsev-Kraichnan model for $2.5D$ flows. 
%
The simplicity of the model allowed us to examine analytically and in detail various limits of the system.
In particular we were able to examine the dynamo properties of the system when the system is close to 
certain classes of flows that dynamo action is `forbidden' by the Zheldovich anti-dynamo theorem.
In particular our results showed that the limits $k_z\to 0$ and $D_r\to 0$ 
(that correspond to 2D magnetic fields and 2D velocity fields respectably) do not commute with $Rm\to\infty$ limit.
This implies that the large $Rm$ results are valid provided that $Rm \gg 1/D_r$, and $Rm\gg k_0/k_z$ and not for the 
exactly 2D case.

Our analysis also allowed us to predict the functional form of the energy spectra of the unstable dynamo modes.  
Two power law behaviours were predicted.
In the range of wavenumbes $k_0\ll k \ll k_z$ the energy spectra 
satisfy $E^{B}_{_{2D}} \propto k^1$ and $E^{B}_{_{Z}}\propto k^3$ 
while in  the range $k_z \ll k \ll k_d$ the spectra satisfy $E^{B}_{_{2D}} \propto E^{B}_{_{Z}}\propto k^0$.
A summary of this behavior is depicted in figure \ref{fig:Plotsmade}. 
These predictions are new and can not be obtained simply by dimensional analysis.

Finally we compared the theoretical results to direct numerical simulation of 
homogeneous, delta-correlated, Gaussian distributed flow and freely evolving flows based on the Navier-Stokes equations.
%
In both the cases the growth rate curves matched qualitatively with the model 
and the magnetic field spectra are in agreement with the theoretical predicted power laws.
This gives support in the relevance of these results to more realistic flows that might occur in nature.

Our study was limited only for a smooth flows. 
An interesting extension would be to study the dynamo instability driven by rough flows that resembles the turbulent flow under fast rotation. 
For the rough flows the holder exponent $\zeta$ for the second order correlation function of the velocity field should take into account the 
Kolmogorov spectra of  $2D$-turbulence. This leads to very interesting possibilities.
For scales smaller than the forcing scale the two-dimensional velocity field ${\bf u}_{_{2D}}$ 
forms a $k^{-3}$ energy spectrum and would continue to follow the $r^2$ scaling for $g_{_{2D}}$.
However the vertical velocity field that is advected like a passive scalar and has a spectrum proportional to $k^{-1}$
would have $g_{_{Z}}\propto r^{0}$ scaling with possible logarithmic corrections. 
For scales larger than the forcing scale an inverse energy cascade develops with  a Kolmogorov energy spectrum $k^{-5/3}$
for ${\bf u}_{_{2D}}$ while $u_z$ reaches a thermalized distribution $k^1$. 
This implies that the correlation function $g_{_{2D}}$ will follow a $r^{2/3}$ 
scaling while the vertical scales will have a much shallower scaling.
We plan to address these possibilities in our future work.

\acknowledgements

The authors would like to thank the group of S. Fauve for their very useful comments and fruitful discussions. 
K. S. would like to thank R. V. K. Chakravarthy for his help in resolving the eigenvalue problem. 
The present work benefited from the computational support of the HPC resources of 
GENCI-TGCC-CURIE \& GENCI-CINES-JADE (Project No. x2014056421 \& No. x2015056421)
and MesoPSL financed by the Region Ile de France and the project EquipMeso (reference ANR-10-EQPX-29-01) where the numerical simulations have been performed.

\appendix
\section{Derivation of the equations} \label{Appendix:One}
In order to derive the equations \ref{eqn:magcorreqns} we follow a procedure similar to the one mentioned in \cite{schekochihin2002spectra}. We start with the index form of the induction equation \ref{eqn:induction} written as,
\begin{align}
\partial_t b^i = b^m \partial^m u^i - u^m \partial^m b^i - i\, k_z \, u_z b^i + \eta \left( \partial^k \partial^k - k_z^2 \right) b^i.
\end{align}
where $\partial^i$ denotes the derivative with respect to the coordinate $x^i$. Next we write the equation for the magnetic correlation function $H^{ij} \left( {\bf r} \right) = \left\langle \left( b^i \left( {\bf x + r}\right) \right)^{\dagger} b^j \left( {\bf x} \right) \right\rangle$, which reads as,
\begin{align}
\partial_t H^{ij} - 2 \eta \Big( \Delta - k_z^2 \Big) H^{ij} & = \partial_k \big[ C^{ikj}\left( {\bf r}, t\right) - C^{kij}\left( {\bf r}, t\right) - \left( C^{jki}\left( - {\bf r}, t\right) \right)^{\dagger} + \left( C^{kji}\left( - {\bf r}, t\right) \right)^{\dagger} \big]  \nonumber \\
& + i k_z \left[ C^{3ij} \left( {\bf r}, t \right) - \left( C^{3ji} \left( -{\bf r}, t \right) \right)^{\dagger} - C^{i3j}\left( {\bf r}, t\right) + \left( C^{j3i} \left( -{\bf r}, t\right) \right)^{\dagger} \right] \label{eqn:Hfull}
\end{align}
where the quantity $C^{kij}$ is the triple product average defined as  $C^{kij}\left( {\bf r}, t\right) = \left\langle u^k \left({\bf x + r}, t\right) \left( b^i \left( {\bf x + r}, t\right) \right)^{\dagger} b^j \left( {\bf x}, t\right) \right\rangle$. This triple product average can be simplified using the Furutsu-Novikov theorem which can be written as,
\begin{align}
C^{kij}\left( {\bf r}, t\right) &=  \left\langle u^k\left( {\bf x + r}, t \right) \left( b^i\left( {\bf x + r}, t \right) \right)^{\dagger} b^j \left({\bf x}, t\right) \right\rangle \nonumber \\
                                &=  \int dx' dt' \left\langle u^k\left( {\bf x + r}, t\right) u^m\left( {\bf x'}, t'  \right) \right\rangle 
                  \left\langle \frac{\delta\left( \left( b^i \left( {\bf x + r}, t \right) \right)^{\dagger} b^j \left( {\bf x}, t\right) \right) }{\delta u^m\left( {\bf x'}, t' \right) }\right\rangle.
\end{align}
The above expression can be simplified by using the delta-correlation property of the velocity correlator. The functional derivative can be simplified by taking the functional derivative of the governing equation of the two point magnetic correlation function $\left( b^i \right)^{\dagger} b^j$. Integrating it with respect to time and taking the statistical average we end up with the following,
\begin{align}
C^{kij}\left( {\bf r}, t\right) = \frac{1}{2} \bigg\{ \Big( & g^{kl} \left( {\bf r}, t \right) - g^{kl} \left( {\bf 0}, t \right) \Big) H^{ij}_{,l} \left( {\bf r}, t \right) - g^{kj}_{,l} \left( {\bf r}, t \right) H^{il} \left( {\bf r}, t \right) - g^{ki}_{,l} \left( {\bf 0}, t \right) H^{lj} \left( {\bf r}, t \right) \nonumber \\
+ & i k_z H^{ij}\left( {\bf r}, t \right) \Big( g^{k3} \left( {\bf 0}, t \right) - g^{k3} \left( {\bf r}, t \right) \Big)  \bigg\} \label{eqn:threept}
\end{align}
We mention here that the Furutsu-Novikov theorem follows the Stratanovich interpretation of the noise as compared to Ito. Substituting the last expression for the triple point averages into the equation \ref{eqn:threept} and after some long but trivial calculation we can find the equation for $H^{ij}\left( r \right)$. 

	Now given the equation for $H^{ij}$ that can be obtained from both \ref{eqn:Hfull} and \ref{eqn:threept}, we look at constructing the equations for scalar functions of $H^{ij}$. The procedure to express the tensor $H^{ij}$ in terms of the possible scalar functions is mentioned in \cite{oughton1997general}. It can then be shown that the correlation tensor $H^{ij}$ has the general form written out in equation \ref{eqn:genformH}. We mention here that only the mirror symmetric part of the correlation function $H^{ij}$ is important in the discussion. This is because the helical part of the magnetic field is not coupled to the governing equations of the nonhelical part. One simple way to see this is to take the equation \ref{eqn:Hfull}, now we use the form of $C^{kij}$ from \ref{eqn:threept}. If we look at an equation governing the proper scalar function in $H^{ij}$, it can be made up of two kinds of terms. One form of the term is a product of two proper scalar functions, more precisely a product of one proper scalar function in $g^{ij}$ and one in $H^{ij}$. The other way is to construct it using the product of two pseudo scalar functions, one pseudo scalar function in $g^{ij}$ and the other from $H^{ij}$. Since there are no pseudo scalar functions in $g^{ij}$ the pseudo scalar functions in $H^{ij}$ do not enter the governing equations of the proper scalar functions in $H^{ij}$. Hence we consider the magnetic correlation function $H^{ij}$ made of only the proper scalar terms, $H_{_{LL}}, H_{_{NN}}, H_{_{Z}}, H_c$. Due to the solenoidal condition we stick with two of these quantities $H_{_{LL}}, H_c$ and their governing equation derived using equations \ref{eqn:Hfull}, \ref{eqn:threept} is mentioned in equation \ref{eqn:magcorreqns}.

\section{Asymptotic forms for correlation functions} \label{Appendix:Two}

The small and large $r$ forms for the correlation functions $h_{_{LL}} \left( r \right)$ and ${h}_c\left( r \right)$ can be obtained from their governing equations \ref{eqn:magcorreqns}. For the case of finite $Rm$ the small $r$ expression reads as, 
\begin{align}
h_{_{LL}} \left( r \right) = & a_0 - \frac{(\gamma + 2 \eta k_z^2 - 8 D_1) a_0 - 4 \eta k_z b_1}{16 \eta} r^2 + O \left( r^4 \right) \\
{h}_c \left( r \right) = & b_1 r - \frac{(\gamma + 2 \eta k_z^2) b_1 - 2 k_z D_2 a_0}{16 \eta} r^3 + O \left( r^5 \right)
\end{align}
here $a_0, b_1$ are constants. For the large $r$ behaviour we have,
\begin{align}
h_{_{LL}} \left( r \right) \sim e^{-\sqrt{\gamma/{2\eta}+k_z^2} \;\; r} \\
{h}_c \left( r \right) \sim e^{-\sqrt{\gamma/{2\eta}+k_z^2} \;\; r}
\end{align}
The case for the $Rm \rightarrow \infty$, we have the rescaled $\tilde{r} = r k_d$ and $\tilde{k}_z = k_z/k_d$. The small $\tilde{r}$ behaviour of the functions $h_{_{LL}} \left( \tilde{r} \right), {h}_c \left( \tilde{r} \right)$ obtained from equation \ref{eqn:magcorrinf} are,
\begin{align}
h_{_{LL}} \left( \tilde{r} \right) = \tilde{a}_0 - \frac{(\gamma+2 \tilde{k}_z^2 -8) \tilde{a}_0 - 4 \tilde{k}_z \tilde{b}_1}{16} \tilde{r}^2 + O(\tilde{r}^4) \\
{h}_c \left( \tilde{r} \right) = \tilde{b}_1 \tilde{r} - \frac{(\gamma+2 \tilde{k}_z^2) \tilde{b}_1 - 2 \tilde{k}_z D_r \tilde{a}_0}{16} \tilde{r}^3 + O( \tilde{r}^5)
\end{align}
where $\tilde{a}_0, \tilde{b}_1$ are constants. For the large $r$ we have,
\begin{align}
h_{_{LL}} \left( \tilde{r} \right) \sim e^{-\sqrt{D_r} \, \tilde{k}_z \tilde{r}} \\
{h}_c \left( \tilde{r} \right) \sim e^{-\sqrt{D_r} \, \tilde{k}_z \tilde{r}}
\end{align}

\section{Spectra of the eigenmode} \label{Appendix:Three}

From the asymptotics we can calculate the power laws of the isotropic spectra of the eigenmode. Using the asymptotic expression from the previous section we reconstruct the following form for the correlation functions $h_{_{LL}}(r), {h}_c(r)$.
\begin{align}
h_{_{LL}} \left( r \right) = & e^{-\sqrt{D_r} k_z r} \sum_{n=0}^{\infty} h_n r^{2n} \\
{h}_{c} \left( r \right) = & e^{-\sqrt{D_r} k_z r} \sum_{n=0}^{\infty} g_n r^{2n}. \label{eqn:correlmodel1}
\end{align}
Now we look for the behaviour of $E^{B}_{_{2D}}(k), E^{B}_{_{Z}} (k)$ for $k \ll k_z$ and $k \gg k_d$. The intermediate range of scales when there is sufficient scale separation between $k_z$ and $k_d$ will be dealt using matched asymptotics. The details of the calculation and the resulting scaling in this intermediate range are mentioned in the following appendix section (see Appendix \ref{Appendix:Four}). 
Using the expression \ref{eqn:correlmodel1} we can obtain an expression for $E^{B}_{_{2D}} \left( k \right), E^{B}_{_{Z}} \left( k \right)$ in the small $k \ll k_z$ limit,
\begin{align}
E^{B}_{_{2D}} \left( k \right) = & c_1 \frac{k}{k_z} + O(k^3) \\
E^{B}_{_{Z}} \left( k \right) = & c_2 \frac{k^3}{k_z^3} + O(k^3)
\end{align}
where $c_1$ and $c_2$ are some constants that are independent of $k$. For scales larger than the dissipative scales $k \gg k_d$ we need to look at the small $r$ behaviour for $h_{_{2D}}, h_{_{Z}}$. We use the steepest descent method for the correlation functions in equation \ref{eqn:correlmodel1} and obtain the following,
\begin{align}
E^{B}_{_{2D}} \left( k \right) = & e^{-k/k_d} \Big( \tilde{c_1} + O(k^{-3/2}) \Big) \\
E^{B}_{_{Z}} \left( k \right) = & e^{-k/k_d} \Big( \tilde{c_2} + O(k^{-3/2}) \Big)
\end{align}
where $\tilde{c}_1$ and $\tilde{c}_2$ are some constants independent of $k$. These behaviour are well captured in the results from the eigenvalue solver (see figures \ref{fig:Spectras}).

\section{Matched Asymptotics} \label{Appendix:Four}

We are interested in finding the behaviour of the functions $E^{B}_{_{2D}}, E^{B}_{_{Z}}$ in the intermediate region $k_z \ll k \ll k_d$. In this process we would would like to find the value of $\gamma$ in the limit of $k_z \ll k_d$. From the numerics we can see that the value of $\gamma$ is $3$ in the limit of small $k_z$ and independent of the value of $D_r$, see figures \ref{fig:Growth1}, \ref{fig:LargeSmallDr}. We are interested in the limit $Rm \rightarrow \infty$ the governing equations are given by equation \ref{eqn:magcorrinf}. Since the equation is rescaled with $k_d$ the small parameter now is $\tilde{k}_z \ll 1$. The idea here is to find the inner solution of the equation by expanding in terms of powers of $\tilde{k}_z$ the equation \ref{eqn:magcorrinf}. Then we compute the outer solution by rescaling the variable $\tilde{r}$ to $\hat{r} = \sqrt{D_r} k_z \tilde{r}$. This rescaling would then provide us with a new set of equations for the outer solution. The behaviour of the inner solution is valid in the region $\tilde{r} \ll 1$ while the outer solution is valid in the region $\tilde{r} \gg 1/\tilde{k}_z$. The matching will take place in the intermediate range of scales, to get the exponents and the eigenvalue $\gamma$.

\subsection{Inner solution}

We do asymptotics for $\tilde{k}_z \ll 1$ with $h_{_{LL}} = H_0 + \tilde{k}_z^2 H_1 + \cdots$ and ${h}_c = D_r \tilde{k}_z \left( G_0 + \tilde{k}_z^2 G_1 + \cdot \right)$, the equation for zeroth order in $k_z$ satisfy,

\begin{align}
\tilde{r}^2 \Big[ H''_0 + 7 \frac{H'_0}{\tilde{r}} - \left( \gamma - 8 \right) \frac{H_0}{\tilde{r}^2} \Big] + \Big[ 2 H''_0 + 6 \frac{H'_0}{\tilde{r}} \Big] = & 0 \\
\tilde{r}^2 \Big[ G''_0 + \frac{G'_0}{\tilde{r}} - \left( \gamma + 1 \right) \frac{G_0}{\tilde{r}^2} \Big] + \Big[ 2 G''_0 + 2 \frac{G'_0}{\tilde{r}} - 2 \frac{G_0}{\tilde{r}^2} \Big] = & 2 \tilde{r} H
\end{align}
Now we write the homogeneous solution to the equations using hypergeometric functions ${}_2F_1$ defined as ${}_2F_1 (a,b,c,d) = \Gamma(c)/(\Gamma(b) \Gamma(c-b)) \int_0^1 t^b (1-t)^{c-b-1}/(1-tz)^{a} dt$,
\begin{align}
H_0\left( \tilde{r} \right) = C_1 \;\; {}_2F_1 \Big[ \frac{3}{2} - \frac{\sqrt{1 + \gamma}}{2},\frac{3}{2} + \frac{\sqrt{1 + \gamma}}{2}, 2; -\frac{\tilde{r}^2}{2} \Big] \\
G_{0H}\left( \tilde{r} \right) = C_2 \;\; \tilde{r} \;\; {}_2F_1 \Big[ \frac{1}{2} - \frac{\sqrt{1+\gamma}}{2},\frac{1}{2} + \frac{\sqrt{1+\gamma}}{2}, 2; -\frac{\tilde{r}^2}{2} \Big]
\end{align}
where $G_0=G_{0H}+ G_{0I}$ with
$G_{0H}$ the homogeneous solution and 
$G_{0I}$ the inhomogeneous solution. $G_{0I}$ can be found and expressed in terms of integrals using the Wronskian.  
The asymptotics for large $\tilde{r}$ is found out to be,
\begin{align}
G_{0I} \left( \tilde{r} \right) = C_1 \frac{1}{\left( \gamma + 1 \right) } \tilde{r}^{-2 \sqrt{\gamma+1}}
\end{align}

\subsection{Outer solution}

For the large $\tilde{r}$ limit we could rescale $\hat{r} \rightarrow \tilde{k}_z \, \tilde{r}$ but in order to get rid of the dependence on $D_r$ at the lowest order we do the following rescaling, $\hat{r} \rightarrow \sqrt{D_r} \tilde{k}_z \, \tilde{r}$. This ends up with the following set of equations, 
\begin{align}
\gamma h_{_{LL}} - \Big( 2 D_r \tilde{k}_z^2 + \hat{r}^2 \Big) \Big[ h_{_{LL}}'' + 3 \frac{h_{_{LL}}}{\hat{r}} \Big] + \Big( 2 \tilde{k}_z^2 + \hat{r}^2 \Big) h_{_{LL}} & = 8 h_{_{LL}} \nonumber \\
+ 4 \hat{r} h_{_{LL}}' + 8 \frac{\hat{r}}{\sqrt{D_r}} {h}_c + \frac{4}{\hat{r}} \sqrt{D_r} & \tilde{k}_z^2 {h}_c \\
\gamma {h}_c - \Big( 2 D_r \tilde{k}_z^2 + \hat{r}^2 \Big) \Big[ {h}_c'' + \frac{{h}_c'}{\hat{r}} - \frac{{h}_c}{\hat{r}^2} \Big] + \Big( 2 \tilde{k}_z^2 + \hat{r}^2 \Big) {h}_c & = 2 \hat{r} \sqrt{D_r} h_{_{LL}}
\end{align}

Since $\tilde{k}_z \ll 1$ we can again expand the quantities $H_{_{LL}} \left( \hat{r} \right), {H}_c \left( \hat{r} \right)$ in powers of $\tilde{k}_z$,
\begin{align}
h_{_{LL}} = \Big[ \hat{H}_0 \left( \hat{r}, \gamma \right) + \tilde{k}_z^2 \hat{H}_1 \left( \hat{r}, \gamma, \tilde{k}_z, \sqrt{D_r} \right) + \tilde{k}_z^4 \hat{H}_2 \left( \hat{r}, \gamma, \sqrt{D_r} \right) + \cdots \Big] \\
{h}_c = \sqrt{D_r} \Big[ \hat{G}_0 \left( \hat{r}, \gamma \right) + \tilde{k}_z^2 \hat{G}_1 \left( \hat{r}, \gamma, \sqrt{D_r} \right) + \tilde{k}_z^4 \hat{G}_2 \left( \hat{r}, \gamma, \sqrt{D_r} \right) + \cdots \Big]
\end{align}
With this expansion the equation at the leading order becomes independent of $D_r$ with the assumptions being $D_r \tilde{k}_z^2 \ll 1, \tilde{k}_z^2 \ll 1$. The leading order equations are,
\begin{align}
\left( \gamma - 8 \right) \hat{H}_0 - \hat{r}^2 \Big[ \hat{H}''_0 + 3 \frac{\hat{H}'_0}{\hat{r}} \Big] + \hat{r}^2 \hat{H}_0 - 4 \hat{r} \hat{H}'_0 & = 8 \hat{r} \hat{G}_0 \\
\gamma \hat{G}_0 - \hat{r}^2 \Big[ \hat{G}''_0 + \frac{\hat{G}'_0}{\hat{r}} - \frac{\hat{G}_0}{\hat{r}^2} \Big] + \hat{r}^2 \hat{G}_0 & = 2 \hat{r} \hat{H}_0
\end{align}

The small $\hat{r}$ behaviour of the functions $\hat{H}_0, \hat{G}_0$ can be obtained by expanding in small $\hat{r}$
By direct substitution it can be shown that a simple power law expansion fails for any value of $\gamma$ and the expansion for small $\hat{r}$ contains logarithmic corrections. 

\subsection{Matching}

We follow the rescale the inner and outer variable to match the solutions at an intermediate range. 
The large $r$ form for the inner solution reads like, 
\begin{align}
H_0 \left( r \right) & =  r^{-\sqrt{1+\gamma}} \Big[ f_1 \left( \gamma \right) \frac{1}{r^3} + f_2 \left( \gamma \right) \frac{1}{r^5} + O{\frac{1}{r^7}} \Big] + r^{\sqrt{1+\gamma}} \Big[ m_1 \left( \gamma \right) \frac{1}{r^3} + m_2 \left( \gamma \right) \frac{1}{r^5} + O{\frac{1}{r^7}} \Big] \\
G_0 \left( r \right) & = r^{1-\sqrt{1+\gamma}} \Big[ \tilde{f}_1 \left( \gamma \right) \frac{1}{r} + \tilde{f}_2 \left( \gamma \right) \frac{1}{r^3} + O{\frac{1}{r^5}} \Big] + r^{1+\sqrt{1+\gamma}} \Big[ \tilde{m}_1 \left( \gamma \right) \frac{1}{r} + \tilde{m}_2 \left( \gamma \right) \frac{1}{r^3} + O{\frac{1}{r^5}} \Big]
\end{align}
for $\gamma \ne 3$. For  $\gamma = 3$ the coefficients $f_i,m_i$ and $\tilde{f}_i,\tilde{m}_i$ diverge. 
In this case the expansion involves logarithmic corrections to the power laws.
A successful matching with the outer solution (that also includes logarithmic corrections) becomes only possible 
for $\gamma=3$. The power law behaviours then for the correlation functions are then a direct consequence of this eigenvalue
and the properties of the hypergeometric functions.
Thus in the intermediate region $r_d \ll r \ll 1/k_z$ the solution has the exponents $h_{_{LL}} \sim r^{-1}, h_c \sim \sqrt{D_r}k_z r^{0}$. Using Wiener-Khintchine we can find the corresponding behaviour in the spectral space to be, $E_{_{2D}}^B \sim k^0 + \frac{1}{2} \sqrt{D_r} k_z/k^2$ and $E_{_{Z}}^B \left( k \right) \sim \frac{1}{2} \sqrt{D_r} k_z k^0$.

\bibliographystyle{jfm}
\bibliography{refs}

\end{document}